\documentclass[fleqn,usenatbib,useAMS]{mnras}

\usepackage{graphicx}	
\usepackage{amsmath}	
\usepackage{amssymb}	
\usepackage{multicol}        
\usepackage{multirow}
\usepackage{bm}		
\usepackage{pdflscape}	
\usepackage{subfigure}
\usepackage{color, colortbl}
\usepackage{siunitx}

\usepackage{xspace}
\usepackage[capitalize]{cleveref}
\usepackage[dvipsnames]{xcolor}

\usepackage[T1]{fontenc}
\usepackage{ae,aecompl}
\usepackage{newtxtext,newtxmath}

\definecolor{Gray}{gray}{0.9}

\newcommand{\prx}{Phys. Rev. X}

\newcommand{\bilbyMCMC}{\textsc{Bilby-MCMC}\xspace}
\newcommand{\dynesty}{\textsc{dynesty}\xspace}
\newcommand{\bilby}{\textsc{Bilby}\xspace}
\newcommand{\pycbc}{\textsc{PyCBC-Inference}\xspace}
\newcommand{\LALInference}{\textsc{LALInference}\xspace}
\newcommand{\LALInferenceNest}{\textsc{LALInference-Nest}\xspace}
\newcommand{\LALInferenceMCMC}{\textsc{LALInference-MCMC}\xspace}
\newcommand{\emcee}{\textsc{emcee}\xspace}
\newcommand{\ptemcee}{\textsc{ptemcee}\xspace}
\newcommand{\kombine}{\textsc{kombine}\xspace}
\newcommand{\Bambi}{\textsc{Bambi}\xspace}
\newcommand{\scipy}{\textsc{scipy}\xspace}
\newcommand{\sklearn}{\textsc{sklearn}\xspace}

\newcommand{\pvtwonrt}{\texttt{IMRPhenomPv2\_NRTidal}}

\newcommand{\nburn}{\ensuremath{n_{\rm burn}}\xspace}
\newcommand{\nlikelihood}{\ensuremath{n_{\ell}}\xspace}
\newcommand{\rburn}{\ensuremath{r_{\rm burn}}\xspace}
\newcommand{\ntemps}{\ensuremath{n_{\rm temps}}\xspace}
\newcommand{\nensemble}{\ensuremath{n_{\rm ens}}\xspace}
\newcommand{\nact}{\ensuremath{n_{\rm act}}\xspace}
\newcommand{\nlive}{\ensuremath{n_{\rm live}}\xspace}
\newcommand{\ndim}{\ensuremath{n_{\rm dim}}\xspace}

\newcommand{\tlike}{\ensuremath{t_{\ell}}\xspace}
\newcommand{\ttotal}{\ensuremath{T}\xspace}
\newcommand{\ncores}{\ensuremath{n_{\rm cores}}\xspace}
\newcommand{\Lone}{\ensuremath{L_{1}}\xspace}
\newcommand{\Hastings}{\ensuremath{\mathcal{H}}\xspace}
\newcommand{\nsamples}{\ensuremath{n_{\rm samples}}}
\newcommand{\neffsamples}{\ensuremath{n_{\rm samples}^{\rm eff}}}
\newcommand{\likelihood}{\mathcal{L}}
\newcommand{\Z}{\mathcal{Z}}
\newcommand{\lnZ}{\ln\mathcal{Z}}

\newcommand{\prior}{\pi}
\newcommand{\data}{d}
\newcommand{\model}{M}

\newcommand{\massratio}{\ensuremath{q}}
\newcommand{\mchirp}{\ensuremath{\mathcal{M}}}

\newcommand{\thetajn}{\ensuremath{\theta_{\rm JN}}}

\newcommand{\vertlabel}[2]{\multirow{#1}{*}{\rotatebox[origin=c]{90}{#2}}}

\title[Bilby-MCMC]{Bilby-MCMC: An MCMC sampler for gravitational-wave inference}

\author[G.\,Ashton \& C.\,Talbot]{
G.\,Ashton$^{1}$
\thanks{Contact e-mail: \href{gregory.ashton@ligo.org}{gregory.ashton@ligo.org}}
and C.\,Talbot$^{2}$
\\
$^{1}$Department of Physics, Royal Holloway, University of London, TW20 0EX, United Kingdom\\
$^{2}$LIGO Laboratory, California Institute of Technology, Pasadena, CA 91125, USA
}

\date{}

\pubyear{2021}

\begin{document}
\label{firstpage}
\pagerange{\pageref{firstpage}--\pageref{lastpage}}
\maketitle

\begin{abstract}
We introduce \bilbyMCMC, a Markov-Chain Monte-Carlo sampling algorithm tuned for the analysis of gravitational waves from merging compact objects.
\bilbyMCMC provides a parallel-tempered ensemble Metropolis-Hastings sampler with access to a block-updating proposal library including problem-specific and machine learning proposals.
We demonstrate that learning proposals can produce over a 10-fold improvement in efficiency by reducing the autocorrelation time.
Using a variety of standard and problem-specific tests, we validate the ability of the \bilbyMCMC sampler to produce independent posterior samples and estimate the Bayesian evidence.
Compared to the widely-used \dynesty nested sampling algorithm, \bilbyMCMC is less efficient in producing independent posterior samples and less accurate 
in its estimation of the evidence.
However, we find that posterior samples drawn from the \bilbyMCMC sampler are more robust: never failing to pass our validation tests.
Meanwhile, the \dynesty sampler fails the difficult-to-sample Rosenbrock likelihood test, over constraining the posterior.
For CBC problems, this highlights the importance of cross-sampler comparisons to ensure results are robust to sampling error.
Finally, \bilbyMCMC can be embarrassingly and asynchronously parallelised making it highly suitable for reducing the analysis wall-time using a High Throughput Computing environment.
\bilbyMCMC may be a useful tool for the rapid and robust analysis of gravitational-wave signals during the advanced detector era and we expect it to have utility throughout astrophysics.
\end{abstract}


\section{Introduction}
\label{sec:introduction}

Gravitational-wave astronomy has enabled the first measurements of masses of merging binary black holes~\citep{abbott16_gw150914_pe}, new constraints on the equation of state of nuclear matter~\citep{abbott18_GW170817_NS_parameters}, and offers a new opportunity to measure the expansion rate of the Universe  \citep{abbott17_gw170817_Hubble} and break the existing measurement tension~\citep{feeney2019}.
In the coming years, the Advanced-LIGO \citep{2015CQGra..32g4001L}, Virgo \citep{2015CQGra..32b4001A}, and KAGRA \citep{kagra} detectors are expected to start a fourth observing run which will see the rate of observed signals from binary mergers increase by up to an order of magnitude.
This brings with it challenges in data analysis: we need software that is rapid, reliable, and can take advantage of available large-scale computing.

Many of the science goals of gravitational-wave astronomy rely on the ability to robustly draw samples from the posterior distribution $p(\theta | \data)$, where $\theta$ are the model parameters and $\data$ is the data, and estimate the Bayesian evidence $\Z$. (For an introduction to Bayesian inference for gravitational-wave astronomy, see, e.g., \citet{thrane2019}).
For Compact Binary Coalescence (CBC) signals, the posterior distribution is highly non-Gaussian with complicated correlations.
Because of the complicated structure of the posterior, stochastic sampling is one of the only viable processes which can robustly generate these quantities for the full complexity of the model (though see \citet{green2020, gabbard2019} for machine-learning based approaches and
\citet{2015PhRvD..92b3002P, 2018arXiv180510457L} for iterative-fitting based approaches).

Stochastic sampling algorithms to analyse CBC signals have predominately applied either a Markov-Chain Monte-Carlo (MCMC) approach \citep{1970Bimka..57...97H, metropolis1953equation}, as introduced by \citet{christensen1998} or Nested Sampling \citep{skilling2006}, as introduced by \cite{veitch08}.
The dominant software used since the first observing run (O1) has been
\LALInference \citep{Veitch2015}, which provides three independent stochastic samplers: two Nested Sampling algorithms, \LALInferenceNest and \Bambi \citep{Bambi}) and a Metropolis-Hastings MCMC (\LALInferenceMCMC) algorithm.
\citet{Veitch2015} provided a variety of standard analytical test cases to demonstrate the validity of each of the samplers.
But, in the absence of analytical posterior distributions for CBC signals, cross-sampler comparisons, especially between different sampling algorithms, are an important check that results are robust.
That \LALInference offered multiple samplers was critical to its success.

The \LALInference software had been widely used, well-tested, and become a benchmark for other packages developed since.
Since its development, a number of high-quality general-use stochastic sampling software packages are now available.
Modern gravitational-wave data-analysis software has been developed which provide an interface to use these off-the-shelf samplers for gravitational-wave astronomy.
For example, both \bilby~\citep{Bilby1} and \pycbc~\citep{Biwer2019} utilise the \dynesty~\citep{dynesty} Nested Sampling algorithm and the \ptemcee~\citep{vousden2016} MCMC algorithm (amongst others).
In \citet{Bilby2}, a detailed cross-sampler comparison was performed between the \bilby-implementation of the \dynesty sampler and \LALInference and the two where found to agreement to within statistical uncertainties. 
However, such samplers rarely work out of the box and often need some customisation and validation to handle CBC signals.
For example, a study by \citet{Kulkarni:2020htf} compared the \ptemcee and \dynesty samplers and found the \ptemcee sampler unable to produce unbiased results for binary black hole systems. 

The \bilby package provides a modular interface to several stochastic samplers and the ability to implement arbitrary likelihoods and priors.
This flexibility has made \bilby a popular choice across astrophysics.
However, our testing with \bilby has revealed that the implemented MCMC algorithms do not produce results which match those of either the \dynesty or \LALInference packages for CBC-like use cases.
This has a significant future impact: the ability to cross-check between samplers remains a critical test for robustness.
In addition, MCMC-based methods are nearly embarrassingly-parallelisable making them ideal for use in a High Throughput Computing (HTC) environment.\footnote{HTC environments differ from High Performance Computing in that the interconnect speeds between nodes is slow.
This makes HTC environments sub-optimal for algorithms which require regular inter-node communication (e.g, the massively parallel methods explored in \citet{pbilby}).
As discussed later in \cref{sec:parallel}, multiple independent MCMC algorithms can be run which continuously produce independent samples, making them ideal for a HTC environment.}
With these two motivations, we verify an MCMC-based algorithm implemented within \bilby as of version 1.1.3.

We began by looking at off-the-shelf options.
These are preferable as they are well-tested and often actively maintained and improved by the open-source community.
An obvious choice which has demonstrated performance for CBC inference \citep{Biwer2019} is the \ptemcee MCMC sampler, an adaptive parallel-tempered version of the \emcee \citep{emcee} algorithm. (The multimodal posterior distributions inherent to CBC problems necessitate the use of the \ptemcee parallel-tempering approach \citep{gilks1998adaptive, earl2005parallel}).
Both of these algorithms use ensemble-sampling in which \nensemble MCMC chains evolve in tandem, new points are proposed based on the position of other chains in the ensemble (cf. \cref{sec:ensemble}).
However, in testing we found \ptemcee to be inefficient compared to the \dynesty sampler (by up to a factor of 100).
In comparison, the \LALInferenceMCMC sampler has a demonstrated efficiency similar to that of the Nest algorithm \citep{Veitch2015}.
Unlike the off-the-shelf samplers, \LALInferenceMCMC utilises a \emph{proposal distribution} (cf. \cref{sec:bilby-mcmc}) which takes advantage of knowledge about the problem in hand (in our case, CBC signals).
This suggests the need for a parallel-tempered MCMC sampler with access to problem-specific proposals.

In this work, we develop \bilbyMCMC, a from-scratch MCMC algorithm with adaptive parallel tempering and ensemble sampling.
We develop \bilbyMCMC to take advantage of a wide variety of proposal distributions including standard, problem-specific and machine-learning based proposals.
We validate the \bilbyMCMC sampler against the \dynesty nested sampler (as described in \citet{Bilby2}) for its use in both standard validation problems and CBC inference.
In this paper, we discuss the tuning and validation of \bilbyMCMC for efficient inference of CBC signals. However, as part of the \bilby package, \bilbyMCMC can be used as a sampler for any inference problem, and includes access to an interface to define custom problem-specific proposal distributions.

 We introduce the \bilbyMCMC sampler in \cref{sec:bilby-mcmc} then discuss the validation for a set of standardised tests in \cref{sec:validation}.
In \cref{sec:gw}, we apply a set of CBC-specific validation tests and describe the performance before concluding in \cref{sec:conclusion}.

\section{Bilby-MCMC}
\label{sec:bilby-mcmc}
MCMC algorithms generate correlated samples from the target distribution, in our case the posterior distribution, by a sequential stepping process.
We now describe the details of the algorithm relevant to the \bilbyMCMC implementation, for a more thorough introduction to MCMC algorithms in astrophysics we refer the reader to recent reviews \citep{Sanjib2017, Hogg2018}.

We apply the Metropolis-Hastings algorithm \citep{1970Bimka..57...97H, metropolis1953equation} to draw samples from the target density
\begin{align}
    p(\theta | \data) \propto \likelihood(\data | \theta) \pi(\theta)\,
\end{align}
where $\likelihood(\data | \theta)$ is the \emph{likelihood} of the model parameters $\theta$ and $\pi(\theta)$ is the \emph{prior} probability of the model parameters.
Throughout, we assume a fixed model $\model$, though formally we note that both the likelihood and priors are model-dependent [i.e. $\likelihood(\data | \theta)$ is more completely written as $\likelihood(\data | \theta, \model)$ and similarly for the prior].

Given a current sample $\theta$, a proposed sample $\theta'$ is generated from a \emph{proposal distribution} $Q(\theta'| \theta)$.
(We discuss proposal distributions in \cref{sec:proposals}).
We accept the proposed sample $\theta'$ with a probability
\begin{equation}
    \alpha = \textrm{min}\left(1, \frac{Q(\theta| \theta')\;\likelihood(\data|\theta') \;\prior(\theta')}{Q(\theta'| \theta)\;\likelihood(\data| \theta)\;\prior(\theta)} \right)\,,
\label{eqn:mh}
\end{equation}
and append $\theta'$ to a \emph{chain} of samples. If the proposal is rejected, the current sample, $\theta$, is appended to the chain.
We implement the Metropolis Hastings step in practice by drawing a random number $u$ from a uniform distribution on the unit interval, if $\alpha \ge u$ the proposal is accepted, otherwise it is rejected.

We initialize the chain with a random draw from the prior distribution $\prior(\theta)$ and iterate the Metropolis-Hasting algorithm to generate a chain of $m$ samples $\{\theta_i\}$ where $i \in [0, m)$.
Samples in the chain are generally correlated.
Independent samples can be obtained from $\{\theta_i\}$ by selecting a subset of $m/\tau$ samples where $\tau$ is the autocorrelation time (ACT) of the chain.
We select the subset by taking a sample every $\tau$ steps.
We iterate the algorithm until reaching the stopping criteria
\begin{equation}
    n_{\rm samples} \ge \frac{m-\nburn}{\gamma \tau}\,,
    \label{eqn:nsamples}
\end{equation}
where $n_{\rm burn}$ is the number of samples discarded to remove the chain initialization, known as the \emph{burn-in} period, and $\gamma\le1$ is a \emph{thinning} factor (in the python interface, $\gamma$ is \texttt{thin\_by\_nact}).
 
To estimate $\tau$, we use the autocorrelation module provided by the \texttt{emcee v3.0} package~\citep{emcee}. This method improves on the traditional approach by adding an automated function to choose the window (c.f. \citet{Hogg2018} along with the software documentation \url{https://emcee.readthedocs.io/en/stable/tutorials/autocorr/}).

We provide a number of automated approaches to estimate the burn-in period $\nburn$. The primary method is a simple scaling: we discard $\rburn$ autocorrelation times, i.e. $\nburn=\rburn\tau$. By default we use, $\rburn=10$, but this scaling factor can be varied by the user through the \texttt{burn\_in\_nact} option. In addition, proposal methods which violate the assumptions of the MCMC algorithm (e.g., using dynamic tuning to improve convergence), set minimum values for $\nburn$ and the user can also specify $\nburn$ directly if these automated approaches fail.

To produce independent samples (matching the behaviour of the \LALInferenceMCMC sampler), we can set the thinning factor $\gamma=1$.
However, thinning is inefficient in the sense that unthinned samples ($\gamma < 1$) are unbiased and provide greater precision for summary statistics \citep{link2012thinning}. In cases where $\gamma < 1$, we differentiate between the number of samples \cref{eqn:nsamples}, and the effective number of samples $\neffsamples=\gamma\nsamples$. For the standard validation tests in this work, we use $\gamma=1$. For the CBC validation tests, we use $\gamma=1/5$ and $\nsamples\ge 25000$. This ensures a minimum number of 5000 independent samples while smoothing posterior plots and providing more accurate summary statistics.

Having introduced the simple Metropolis-Hastings algorithm, we now turn to the specifics of the \bilbyMCMC implementation which enable it to efficiently perform CBC parameter estimation.
In \cref{sec:proposals}, we define the standard library of proposal distributions then introduce the learning proposals in \cref{sec:learning-proposals} and the gravitational-wave specific proposals in \cref{sec:gw-proposals}.
We note that \bilbyMCMC provides a flexible interface to define and use new proposals. As such, this list is not an exhaustive set of all available proposals.
In \cref{sec:block}, we describe how the proposals are used together in a block-updating sampling approach. 
In \cref{sec:pt}, we describe the extension to a parallel-tempered sampler required to analyse multi-modal distributions then describe how ensemble-sampling is implemented in \cref{sec:ensemble}.
In \cref{sec:efficiency}, we provide a model for the efficiency of the sampler, then introduce a timing model and discuss computational parallelisation in \cref{sec:parallel}.

\subsection{Standard proposal distributions}
\label{sec:proposals}

Broadly speaking, the performance of an MCMC sampler is determined by the ACT of the chains it produces. 
Chains with smaller ACTs taker fewer steps to traverse the target distribution and hence produce more independent samples for a fixed number of MCMC steps (or equivalently, computational cost).
The ACT itself depends on how efficient the proposal distribution $Q(\theta'|\theta)$ is in proposing points which enable the chains to traverse the posterior.

For the Metropolis Hastings algorithm, there are two ways to optimize a stochastic sampler to reduce the ACT.
First, we can choose a parameterisation which reduces the complexity of the parameter space.
If under a transformation $T$, the posterior distribution has a simpler form (e.g, if $T$ maps a Banana-like distribution to a multivariate Gaussian, or softens hard edges), then sampling in $T(\theta)$ rather than $\theta$ is the most straight forward approach to improving the algorithm performance \citep{Hogg2018}.
In \cref{sec:parameterisation}, we discuss the best known parameterisation for CBC signals. 
Second, once the best known parameterisation is chosen, we optimize the choice of proposal distributions.

In this section, we introduce the standard library of proposals implemented in \bilbyMCMC and discuss their performance and utility.
For each proposal, we also provide the \emph{Hastings factor}
\begin{align}
    \Hastings = \frac{Q(\theta| \theta')}{Q(\theta'| \theta)}\,,
\end{align}
which ensures detailed balance is met \citep{1970Bimka..57...97H} and enables unbiased sampling using asymmetric proposals.

\subsubsection{FG: Fixed Gaussian}
\label{sec:fg}
The Fixed Gaussian proposal implemented in \bilbyMCMC is a generalisation of the zero-mean multivariate Gaussian proposal \citep{gelman1996efficient} in which a proposal for the $i^{th}$ parameter is generated from
\begin{equation}
    \theta_i' = \theta_i + \sigma_i w_i \epsilon\,,
    \label{eqn:fixed}
\end{equation}
where $\sigma_i$ is a user-defined scale parameter, $w_i$ is the prior support (if the prior has infinite support, we set $w_i=1$) for $\theta_i$, and $\epsilon$ is a draw from a standard normal distribution.
The introduction of the scaling by the prior support enables some automatic tuning to the anticipated scale of the problem, while the $\sigma_i$ enables the user to define varying length scales for each parameter. Of note, our implementation does not allow the user to change the spatial orientation of the proposal (i.e. through correlations between parameters).
For the Fixed Gaussian proposal, which is symmetric, \Hastings=1.

In practice, the Fixed Gaussian proposals have limited use and require manual tuning (through the choice of $\sigma_i$) to achieve meaningful performance on realistic problems. As such, we do not enable this proposal by default.

\subsubsection{AG: Adaptive Gaussian}
\label{sec:ag}
To circumvent the tuning requirements of a Fixed Gaussian proposal, \citet{haario2001adaptive} introduced the notion of an adaptive proposal which uses past performance of the sampler to drive the sampler to a target acceptance rate.

Such adaptive proposal are non-Markovian and may lead to the generated samples not being representative of the posterior. However, as discussed in \citet{haario2001adaptive} and \citet{Veitch2015} (in the context of CBC signals), if the adaptation rate decays throughout the run or the adaptation is halted sufficiently early in the run, the equilibrium distribution may be sufficiently close to the  posterior. We verify that, to within the statistical uncertainties relevant for typical CBC problems, this is true for our Adaptive Gaussian proposal.

To dynamically adapt the proposal, we use the acceptance ratio:
\begin{equation}
a = \frac{n_{\rm accepted}}{n_{\rm accepted} + n_{\rm rejected}}\,,
\end{equation}
to quantify the current performance \citep{gelman1996efficient}.
If $a \sim 1$, proposals are accepted too often: this suggests slow exploration of the posterior. If $a \ll 1 $, proposals are infrequently accepted: the proposed points tend to jump away from areas of high posterior support.
In both cases, this leads to large autocorrelation times.
For well-tuned proposals (which reduce the ACT relative to poorly-tuned proposals) and under idealised settings, \citet{roberts1997weak} demonstrated that $a \sim 0.23$.
We set this as a target acceptance rate and dynamically adapt the proposal to achieve it.

Our implementation of the Adaptive Gaussian proposal extends \cref{eqn:fixed}:
\begin{equation}
    \theta_i' = \theta_i + s \sigma_i w_i \epsilon
\end{equation}
adding a factor of $s$, the adapting scale parameter. Initially, $s=1$, then on each iteration which the proposal is applied, we update $s$ following~\cite{Veitch2015}:
\begin{align}
    s &\rightarrow s + s_{\gamma} \frac{(1 - a')}{100}\,,
\end{align}
if the previous proposed point was accepted or
\begin{align}
    s &\rightarrow s - s_{\gamma} \frac{a'}{100}\,,
\end{align}
if the previous proposed point was not accepted.
Here $a'=0.234$ is the target acceptance rate and the quantity
\begin{align}
    s_{\gamma} = \left(\frac{N}{n}\right)^{1/5} - 1\,,
\end{align}
is the adaptation decay rate with $n$ the number of points proposed and $N$ a user-specified number of steps after which to stop adapting, the default value is $N=10^5$.
We set a minimum scale $s \ge N^{-1}$.
For the Adaptive Gaussian proposal, which is symmetric, \Hastings=1.

\subsubsection{DE: Differential Evolution}
\label{sec:de}
We implement the Differential Evolution proposal \citep{ter2006, ter2008} as described in \citet{Veitch2015}. Two samples $\theta^a$ and $\theta^b$ are drawn at random from the chain. Then
\begin{equation}
    \theta' = \theta + \gamma(\theta^a - \theta^b)\,,
\end{equation}
where $\gamma$ is chosen randomly from $ \gamma = \left\{1, N(0, 2.38/\sqrt{2 \ndim}\right\}$.
When $\gamma=1$, this acts as a mode-hopping proposal improving the performance of the sampler in multi-modal problems. 
When $\gamma$ is drawn from the normal distribution [as proposed by \citet{ter2006, roberts2001}], the proposed points lie along the line passing through $\theta$ and $\theta'$.
As such, the proposal is well suited to posterior distribution with linear correlations in which the line passing through $\theta$ and $\theta'$ lies along the principle axes.
As with the Adaptive Gaussian proposal, formally this proposal makes the chain non-Markovian. Later, in \cref{sec:validation}, we verify that the equilibrium distribution is statistically identical to the posterior, i.e the posterior is unbiased.
Like the Gaussian proposals, the Differential Evolution proposal is symmetric, such that \Hastings=1.

\subsubsection{PR: Prior Proposal}
\label{sec:pr}
The prior proposal draws $\theta'$ from the prior $\pi(\theta)$.
For well-measured parameters, in which the posterior is much narrower than the prior, this proposal is highly inefficient. However, we find it to be effective when used as part of a block-updating set of proposals applied to poorly measured parameters (e.g., the spin and tidal parameters of the secondary lower-mass object in a CBC inference problem).
It also aids mode-mixing in high temperature chains (see \cref{sec:pt}).
For the Prior proposal, the Hastings factor is $\Hastings=\pi(\theta)/\pi(\theta')$.

\subsubsection{UN: Uniform Proposal}
\label{sec:un}
A simplification of the Prior Proposal, this proposal proposes points uniformly within the prior bounds.
We utilise this proposal as a robust and simpler variant of the Prior Proposal with similar performance. 
For the Uniform proposal, which is symmetric, the Hastings factor is $\Hastings=1$.

\subsection{Machine learning proposal distributions}
\label{sec:learning-proposals}
In \bilbyMCMC, we introduce a class of \emph{learning} proposals which, as we show in \cref{sec:validation}, dramatically decrease the ACT while producing statistically identical posterior distributions.
learning proposals use a random sampling from the past MCMC chain to learn the distribution and then generate new samples.
For all learning proposals, during an initialization stage (during which the MCMC chain has not yet been explored), they fall back to an Adaptive Gaussian proposal.
Once the initialization stage is complete, they sample the MCMC chain, use the samples to fit the proposal distribution, and then this distribution is used to propose new points.
Periodically, the proposal distribution is refitted using new samples from the MCMC chain to circumvent premature learning.
As with the Adaptive Gaussian proposal, the use of the past chain again breaks the Markovian property of the chain, but we verify in \cref{sec:validation} that the resulting posterior remains unbiased using validation tests.

\subsubsection{KD: Gaussian Kernel Density Estimate}
\label{sec:kd}
We fit a Gaussian Kernel Density Estimate (KDE) \citep{kde_1, kde_2} to a random draw of samples from the MCMC chain.
We find this non-parametric multivariate density estimate to be both rapid in learning (typical learning times are fractions of a second) and flexible enough to fit complicated features.
KDE methods have previously been used in the context of CBC inference by the \kombine \citep{kombine} ensemble sampler.

When used to estimate a probability density from a set of samples, KDE methods suffer a subtle dependence on a tuneable ``bandwidth'' parameter and typically over-smooth hard edges and multi-modal distributions.
However, when used as a learning proposal density, these issues only result in loss of efficiency, and do not bias results.
To understand why, consider a parameter with a hard edge (e.g. the lower bound on the spin of a black hole which cannot be negative).
A KDE proposal fitted to a chain will over-smooth the hard edge and propose nonphysical points with negative spin.
However, the MCMC algorithm will never accept these points.
This results in a small loss of efficiency, but no bias.

We utilise the standard implementation of Gaussian KDE in the \scipy \citep{scipy:2020} package with bandwidth estimated using ``Scotts rule'' \citep{scott2015}.
Once the KDE $k(\theta)$ is fitted, proposal samples can be drawn directly and the Hastings factor calculated by $\Hastings=k(\theta)/k(\theta')$.
We find that fitting the KDE takes fractions of a second while the proposal time is negligible compared to typical CBC likelihood evaluation times.

\subsubsection{GM: Gaussian Mixture Model}
\label{sec:gm}
While KDEs smooth a set of samples as a Gaussian centered on each sample, in a Gaussian mixture model (GMM) the density is estimated using a finite number of Gaussian distributions.
As with KDE methods, this model is not good at fitting distributions with hard edges.
The means and covariance matrices of these Gaussian distributions are chosen using an expectation-maximisation algorithm.
We use the \sklearn \citep{sklearn:2012} package to fit the GMM.
In this work, we use 10 components in the mixture.
Fitting the GMM takes slightly more time than fitting the KDE, however, it is typically <\SI{1}{s} and sampling from/evaluating the GMM is faster than sampling from the KDE as there are fewer components.
As with the KDE proposal, the Hastings factor is calculated from $\Hastings=g(\theta)/g(\theta')$ where $g(\theta)$ is the fitted GMM.

\subsubsection{NF: Normalizing Flows}
\label{sec:nf}
The \emph{normalizing flows} class of machine learning algorithms \citep{papamakarios2019normalizing}
learn a bijective map from the target density (the set of training samples drawn from the MCMC sampler) to a latent space, in our case a multivariate Gaussian.
Normalizing flows have previously been used in gravitational-wave astronomy to directly sample the CBC posterior distribution \citep{green2020, 2020arXiv200803312G} and as way to propose new points in a nested sampler \citep{williams2021}.
Following the work of \citet{hoffman2019neutra, moss2020}, we use the \texttt{nflows} package \citep{nflows} which implements the normalizing flows algorithm in \texttt{PyTorch} \citep{pytorch}, to learn the proposal distribution.
We periodically optimize the normalizing flow using the Jensen-Shannon divergence test (cf \cref{sec:js}) between samples drawn from the learnt map and a set of independent validation samples.
Unlike the KDE and GMM proposals, the normalizing flows proposals can take several tens of seconds to minutes to train.
In practice (cf. \cref{sec:validation}), we find the normalizing flows method has a similar performance to the GMM method, but at an increased computational cost. Therefore, we do not utilize it for CBC inference problems.
The Hastings factor is again given by the ratio of the normalizing flow density at the initial and proposed points.

\subsection{Gravitational-wave proposal distributions}
\label{sec:gw-proposals}

We implement the gravitational-wave specific \emph{polarization and phase correlation}, \emph{phase reversal}, and \emph{phase and polarization reversal} proposals as described in \citet{Veitch2015}.
We find these proposals dramatically improve the sampling for analyses which do not utilize analytic marginalization of the binary phase (cf. \cref{sec:marginalization}).
We do not implement the \emph{sky reflection}, \emph{extrinsic-parameter}, and \emph{Gibbs sampling of distance} proposals described in \citet{Veitch2015, raymond2014}.
While we expect these to be general improvements to the algorithm, the use of distance marginalization  (cf. \cref{sec:marginalization}), and our choice of parameterisation (cf. \cref{sec:parameterisation}) diminish the expected utility of these proposals.

\subsection{Block sampling}
\label{sec:block}
Each of the proposal distributions described in the last three sections can update either all parameters in the set of model parameters $\theta$, or only a subset of those parameters.
The \bilbyMCMC sampler is initialised with a list of individual proposals, the subset of $\theta$ which they are to update, and their unnormalised weighting.
We then use the weighting to create a cyclic \emph{proposal cycle}.
At each step of the sampler, the next proposal in the cycle is chosen, a point is proposed and accepted/rejected based on the condition described in \cref{eqn:mh}.
The proposal cycle enables weighted block-updating of proposals and ensures the detailed balance condition is met as described in \citet{Veitch2015}.

For non-CBC inference problems (i.e, the standard tests considered in \cref{sec:validation}), we default to an equal-weighted set of the Adaptive Gaussian, Differential Evolution, Uniform, KDE, GMM, and Normalizing Flow proposals.
Though, this can be customized by users.
For CBC inference problems, we define a proposal cycle described in \cref{tab:gw-proposal-set}.
We arrived at this choice by hand tuning: running analyses on simulated signals and identifying opportunities for improvement.
As such, we do not anticipate that the proposals selected in \cref{tab:gw-proposal-set} are optimal and we expect improvements to be made in the future. Users can modify and extend proposal cycle using the flexible interface.

\begin{table}
    \centering
    \begin{tabular}{l|ll}
         Proposal & $\theta$-subset & weight  \\\hline\hline
         Adaptive Gaussian & all & 10 \\
         Differential Evolution & all & 10 \\ \hline
         Adaptive Gaussian & intrinsic & 10 \\
         Differential Evolution & intrinsic & 10 \\
         KDE & intrinsic & 10 \\
         GMM & intrinsic & 10 \\ \hline
         Differential Evolution & extrinsic & 10 \\
         KDE & extrinsic & 10 \\
         GMM & extrinsic & 10 \\
         Adaptive Gaussian & extrinsic & 5 \\ \hline
         Differential Evolution & mass & 5 \\
         GMM & mass & 5 \\ \hline
         Differential Evolution & spin & 5 \\
         GMM & spin & 5 \\ \hline
         Adaptive Gaussian & measured-spin & 5 \\ \hline
         Differential Evolution & mass ratio and primary spin & 5 \\
         Differential Evolution & tidal deformability & 5 \\
         Prior proposal & tidal deformability & 5 \\
         Phase Reversal & phase & 0.1 \\
         Phase and Polarisation Reversal & phase and polarisation & 0.1 \\
         Correlated Phase/Polarisation & phase and polarisation & 0.1 \\
         Prior & $\psi$, $\phi_{12}$, $\theta_2$, $\Lambda_1$, $\Lambda_2$, $t_j$ & 0.1 \\
    \end{tabular}
    \caption{The gravitational-wave proposals set used in this work.
             For a description of the proposals themselves, see \cref{sec:proposals}.
             In cases where the $\theta$-subset is ``all'', the whole set of $\theta$ is updated.
             Where a subset is listed, see \cref{sec:parameterisation}, only that subset is updated by the proposal.
             The weights are unnormalised and determine the relative frequency of each proposal.
             In the final row of ``Prior'' proposals, each is updated individually, not as a set.
             }
    \label{tab:gw-proposal-set}
\end{table}

\subsection{Parallel-tempering}
\label{sec:pt}
The standard Metropolis Hastings algorithm does not produce estimates of the evidence, and fails when attempting to sample from multi-modal distributions. Parallel-tempering \citep{gilks1998adaptive, earl2005parallel} addresses both of these issues. 

As the name suggests, $\ntemps$ parallel MCMC chains are run. (In practice, these can be updated sequentially, i.e. stepping each chain in turn, or using the parallelisation techniques described later in \cref{sec:parallel}. However, the chains must remain pseudo-synchronised to enable swaps between chains). For the $j^{\rm th}$ chain, the likelihood in \cref{eqn:mh}, is modified:
\begin{equation}
    \likelihood(\data| \theta, M) \rightarrow 
    \likelihood(\data| \theta, M)^{1/T_{j}}
\end{equation}
where $T_{j} \ge 1 $ is the chain ``temperature''.
Note that the ladder of temperatures $\{T_{j}\}$ is ordered $T_{j+1} > T_{j}$.
The $T_0=1$ ``cold'' chain samples from the target posterior distribution. But, for ``hot'' chains with $T_{j} > 1$, the likelihood is flattened out and easier to sample. 

Periodically, swaps are proposed between adjacent chains and accepted with a probability
\begin{align}
    \textrm{min}\left[1, \left(\frac{\likelihood(\data|\theta_m)}{\likelihood(\data|\theta_n)}\right)^{\frac{1}{T_{n}}-\frac{1}{T_{m}}}\right]\,.
\end{align}
These swaps provide a mechanism for the cold temperature chain (which generates posterior samples) to explore multi-modal likelihoods.
We utilize the dynamic temperature adaption methods described in \citet{vousden2016} to
optimize the choice of the temperature ladder $\{T_{j}\}$.
Samples taken during this optimization period are automatically labelled as part of the burn-in epoch.

\subsection{Evidence calculation}
\label{sec:evidence}

In addition to resolving the problem of sampling multimodal distributions, parallel-tempering also enables an estimate to be made of $\lnZ$, the natural logarithm of the Bayesian evidence.
To estimate the evidence in \bilbyMCMC, we implement \emph{thermodynamic integration} \citep{goggans2004using, lartillot2006computing} as described in \citet{littenberg2009, Veitch2015} and the \emph{stepping-stone algorithm}~\citep{Xie2010, russel2019}.
In testing, we verify the findings of \citet{russel2019}: the stepping stone method is superior, producing more accurate results for the same computational cost.
As such, while \bilbyMCMC calculates both methods, we report only the stepping stone evidence throughout this work.

\subsection{Ensemble sampling}
\label{sec:ensemble}
In recent years, ensemble-sampling algorithms have been highly successful in astrophysics (see, e.g. \citet{emcee, vousden2016, kombine}). These algorithms use an ensemble of interacting MCMC samplers.
New points are proposed based on the current distribution of the ensemble of points enabling automatic tuning of the proposals to the target density. That these algorithms self-tune has been paramount to their versatility and use throughout astrophysics.

In \bilbyMCMC, an ensemble of \nensemble chains can be utilised with inter-chain swaps proposed by an ensemble stretch proposal \citep{goodman2010}. In comparison to the \emcee and \ptemcee samplers, \bilbyMCMC is poorly vectorised and does not scale to many hundreds of chains.

If used in conjunction with parallel-tempering, one can either use \ntemps ensembles (for a total of $\ntemps\times\nensemble$ samplers) or with one parallel-tempered chain (for a total of $\ntemps + \nensemble - 1$ samplers). The former configuration mirrors how the \ptemcee sampler operates while the latter configuration may be useful, for example, to provide an estimate of the evidence with a reduced computational cost.
For thermodynamic integration, each set of parallel-tempered chains is used to calculate an estimate of the evidence, then the results are averaged between chains.
In the validation tests described in \cref{sec:validation} and \cref{sec:gw}, we do not utilise the ensemble sampler as it was found to provide no practical improvement in efficiency.

%

\subsection{Efficiency}
\label{sec:efficiency}
Throughout this work, we will quantify and compare the posterior sampling \emph{efficiency} of samplers by the ratio of the number of independent samples to the number of likelihood evaluations:
\begin{align}
    \epsilon = \frac{\neffsamples}{\nlikelihood}\,.
    \label{eqn:efficiency-definition}
\end{align}
For a simple MCMC sampler, the number of steps is equal to the number of likelihood evaluations.
However, in \bilbyMCMC we do not evaluate the likelihood if the proposed sample is outside the prior bounds.
Nevertheless, the number of likelihood evaluations is the relevant weighting as it is the dominant computational operation.

We calculate the efficiency directly for the validation tests in \cref{sec:validation}, but here we first derive an efficiency model.
For a parallel-tempered ensemble sampler with $\ntemps\times\nensemble$ chains where $\nburn=\rburn\tau$ are discarded for burn-in (in practice, there are several alternative methods which can determine $\nburn$ as described in \cref{sec:bilby-mcmc}), the efficiency is
\begin{align}
    \epsilon = \frac{1}{\tau \ntemps(1 - \xi)}\,,
    \label{eqn:efficiency-model}
\end{align}
where
\begin{align}
\xi=\frac{\rburn \nensemble}{\neffsamples}\,,
\label{eqn:burn-in-inefficiency}
\end{align}
is the \emph{burn-in inefficiency}, the fraction of `wasted' samples due to the burn in process.

When configuring the sampler, care should be taken to ensure $\xi \ll 1$ to avoid significant wasted computation.
For example, drawing 1000 independent samples using $\nensemble=1$ and the default $\rburn=10$, the burn-in inefficiency is a reasonable 1\%.
However, if we attempt to use 10 co-evolving ensembles $\nensemble=10$, the burn-in inefficiency also increases to 10\%.
(The same logic equally applies to configurations which combine independent runs as discussed in \cref{sec:parallel}, replacing \nensemble with the number of independent runs and \neffsamples with the number of samples per run.).

Provided $\xi \ll 1$, the efficiency is determined by the ACT $\tau$ and the number of parallel-tempered chains $\ntemps$.
The ACT is a property of the sampling algorithm, which can be reduced using the methods discussed in \cref{sec:proposals}.
Naively, reducing \ntemps appears to improve the posterior sampling efficiency.
However, \ntemps>1 is required to sample from multimodal distributions, calculate the evidence, and can reduce $\tau$.
In \cref{sec:validation} we will demonstrate with a specific example, but as a rough rule of thumb about $\ntemps=8$ is sufficient for the multi-modal posteriors of CBC inference problems and provides a reasonable estimation of the evidence.
However, if a refined estimate of the evidence is required, more temperatures are needed, decreasing the posterior sampling efficiency.

We develop a resampling approach to reclaim some of this lost efficiency.
For the $j^{\rm th}$ chain with temperature $T_j$ we define $\{\theta_{i}\}_{(j)}$ as the set of posterior samples it produces from the tempered posterior distribution:
\begin{align}
    p(\theta | \data) \propto \likelihood(\data | \theta)^{1/T_j}\pi(\theta)
\end{align}
For each sample $\theta$ from the tempered posterior distribution, we calculate a weight
\begin{align}
    w = \likelihood(\data| \theta)^{1-\frac{1}{T_j}}\,
\end{align}
from the ratio of the hot likelihood to the $T=0$ likelihood.
Then, we rejection sample\footnote{We draw $u$ from a uniform distribution on the unit interval, if $u<w$, the sample is accepted, otherwise it is rejected.}~\citep{mackay2003information} the tempered posterior samples resulting in a set of posterior samples from the cold posterior distribution.
For low dimensional problems, we find this produces a modest gain in efficiency at no additional cost.
As an example, analysing a CBC signal using a non-spinning model and using the analytic marginalization of the distance, phase and time (cf. \cref{sec:marginalization}), rejection sampling the hot chains produces a $\sim20$\% efficiency improvement.
However, for fully-precessing CBC problems we find the rejection sampling does not accept any new points (i.e. the efficiency remains unchanged).
In \cref{sec:validation} and \cref{sec:gw}, we do not utilise the rejection sampling method.

In this work, we will compare the efficiency of \bilbyMCMC with that of the \dynesty sampler using the random walk proposal method described in \citet{Bilby2}. 
This proposal method has a tuning parameter \nact which determines the number of
internal MCMC steps to take based on the estimated autocorrelation time.
Following \cite{dynesty} (in which the \dynesty sampler was shown to be more efficient than the \emcee sampler), we calculate the efficiency from \cref{eqn:efficiency-definition}, with $\neffsamples$ calculated from the \emph{effective sample size} as estimated from the nested sampling weights.
We note that this assumes that new points proposed during nested sampling are independent, however this isn't required \citep{salomone_2018} or guaranteed in practice.
If the points are correlated, \neffsamples will significantly overestimate the efficiency of the \dynesty sampler.
To guard against this (and to investigate the potential impact on posterior estimation), for the Rosenbrock and Unimodal Gaussian validation tests, we run the \dynesty sampler with two different values of \nact and verify that the efficiency approximately scales with \nact.
This demonstrates that the \nact value chosen is sufficiently large to generate independent samples and hence that the efficiency of the \dynesty is not overestimated.

\subsection{Timing model and parallelisation}
\label{sec:parallel}

We distinguish two levels of computational parallelisation which can reduce the wall time: \emph{combining independent runs} and \emph{multiprocessing} an individual run. 

Combining independent runs is embarrassingly parallel: we simply repeat $N$ copies of the analysis using an identical data and configuration, but a different random seed.
If each analysis produces $\neffsamples$ independent samples, then in total we end up with $N\neffsamples$.
Such a configuration is ideal for use in a HTC environment and has the added advantage that one can cross-compare between chains.
However, this approach is limited by the increase in burn-in inefficiency (cf. \cref{eqn:burn-in-inefficiency}): each independent run has to burn-in.
This may be worthwhile to decrease the wall-time for important and time-sensitive results.

Before discussing the use of multiprocessing, we introduce a timing model to understand the wall-time required to produce \neffsamples independent samples from the posterior of a single serial run.
For CBC inference problems, the compute-time is determined by $\tlike$, the time required to evaluate the likelihood\footnote{Typically $\tlike$ ranges from a few to many hundreds of milliseconds and is dominated by the cost to evaluate the waveform. Longer-duration signals typically take longer to evaluate. However, when calculating the likelihood during an MCMC analysis, cached waveform evaluations can be used, e.g. when proposing a move only in the extrinsic parameters. The discussion in this section assumes a fixed $\tlike$, resulting in a conservative timing estimate which ignores this potential computational saving.} and the number of likelihood evaluations required. In a serial-processing model, the total time $\ttotal$ can be estimated by
\begin{align}
    \ttotal = \nlikelihood \tlike = \frac{\nsamples \tlike}{\epsilon} \approx 
    \SI{28}{hrs}
    \left(\frac{\neffsamples}{1000}\right)
    \left(\frac{\tlike}{\SI{10}{ms}}\right)
    \left(\frac{\epsilon}{0.01\%}\right)^{-1}\,,
    \label{eqn:serial-timing}
\end{align}
where we have taken a typical efficiency from \cref{sec:BBHA} for an CBC analysis using $\ntemps=8$.

If either $\ntemps>1$ or $\nensemble >1$, \bilbyMCMC can be trivially parallelised leveraging the multi-core processors typically available in modern processors.
We implement this parallelisation using the python standard-library \texttt{multiprocessing} module.
In this model of multiprocessing, there is an overhead cost to transferring the data (i.e. any data products required by the likelihood). For typical CBC problems, this can be as much as a few milliseconds. [We note that the LALInference MCMC sampler \citep{Veitch2015} mitigates this by the use of a distributed computing model with a Message Passing Interface].
This overhead time is comparable to the likelihood evaluation time $\tlike$ and results in imperfect scaling of the timing model \cref{eqn:serial-timing}.
We model this by introducing $m\le1$, a \emph{parallelisation inefficiency} which we will measure empirically.
Then, the timing model for an analysis parallelised over $\ncores$ is
\begin{align}
    \ttotal = \frac{\neffsamples \tlike}{\epsilon} \frac{1}{m\ncores}\,.
    \label{eqn:parallel-timing}
\end{align}
The number of cores should be matched to the number of parallelisable jobs, i.e. $\ncores=\frac{\ntemps \nensemble}{m}$ where $m$ is a non-zero natural number.
If the number of cores is mismatched with the number of parallelisable jobs, i.e. $\ncores > \ntemps\nensemble$, this will always leave one or more cores idle.
When $\ncores = \ntemps\nensemble$, we refer to this as \emph{perfect matching}.

We can measure $m$ empirically by looking at the speed-up for an identical analysis as a function of $\ncores$.
We find that direct parallelisation results in values of $m$ which are as small as (or in the worst case smaller than) $1/\ncores$.
I.e., the parallelisation can be slower than a serial run.
This is because of the substantial data-transfer overhead.
To mitigate the data transfer overhead, in parallel analyses, we transfer data and then take a fixed number, \Lone, of `internal' MCMC steps.
To further improve the efficiency, we do not store these internal steps.
In effect, this pre-thins the MCMC chains by a factor of \Lone.
When \Lone > 1, the ACT and other associated quantities are calculated on the stored chain, but can be re-scaled.

\begin{figure}
    \centering
    \includegraphics[width=0.5\textwidth]{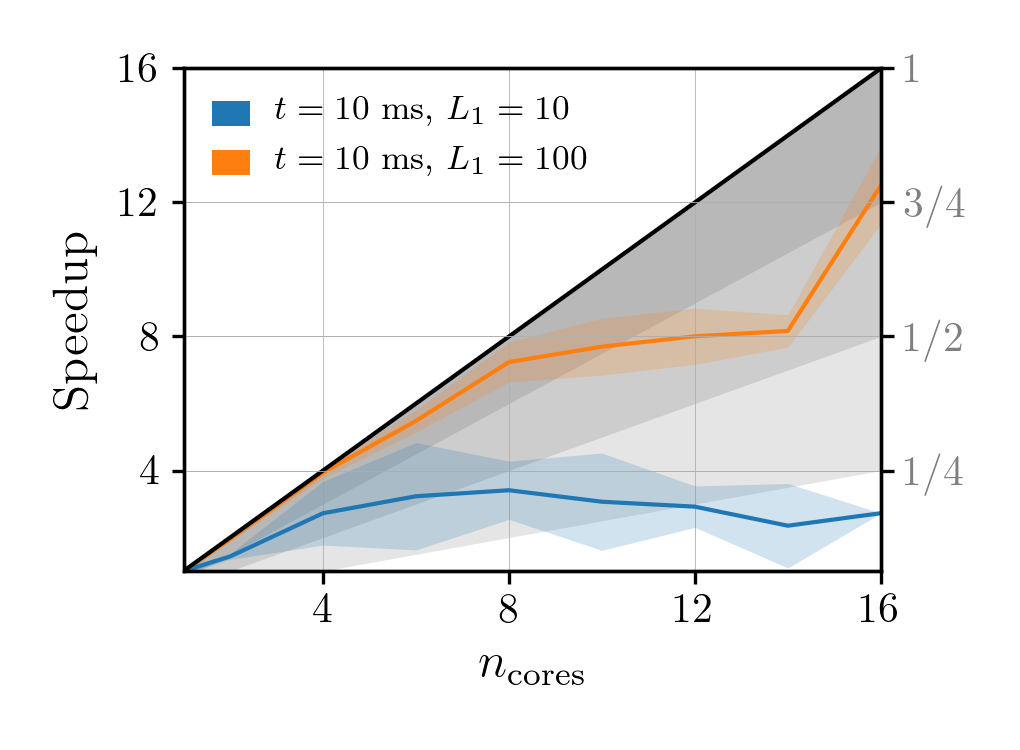}
    \caption{Empirically measured speed-up for a test analysis with $\ntemps=16$ and $\nensemble=1$. Solid lines indicate the mean while shaded region indicates the standard deviation as measured over 3 identical runs. The right-hand axis provides the speed-up factor $m$ at perfect matching.}
    \label{fig:speed-up}
\end{figure}

In \cref{fig:speed-up}, we determine the speed-up factor $m$ for a test case in which $\ntemps=16$ and the per-likelihood evaluation time $\tlike$ is held fixed at \SI{10}{ms}.
We run the experiment twice. First, we use \Lone{=}10, which demonstrates poor parallelisation scaling with an overall speed-up factor of $m{\sim}1/8$ for perfect matching.
Then, we increase \Lone to 100 and see improved scaling with $m{\sim} 3/4$ for perfect matching.
For $\ncores{=}8$ or less, the performance is near-optimal $m \sim 1$.
The marginal improvement in speed up between \ncores of 10, 12, and 14 demonstrates the effect of imperfect matching.

Using the empirically measured $m$ from \cref{fig:speed-up}, if our analysis using $\ntemps=8$, $\nensemble=1$, we can see the rough timing predicted by \cref{eqn:parallel-timing} for perfectly-matched parallelised runs:
\begin{align}
    \ttotal \approx
    \SI{5}{hrs}
    \left(\frac{\neffsamples}{1000}\right)
    \left(\frac{\tlike}{10{\rm ms}}\right)
    \left(\frac{\epsilon}{0.01\%}\right)^{{-}1}
    \left(\frac{m}{0.75}\right)^{{-}1}
    \left(\frac{\ncores}{8}\right)^{{-}1}\,.
    \label{eqn:timing}
\end{align}
It is worth pointed out two caveats to this timing model.
First, while increasing \Lone improves $m$, if \Lone is greater than the typical ACT, this itself introduces a new type of inefficiency (namely an over-thinned chain).
Second, these quantities are not independent.
For example, if one wants to combine a large number of independent runs, each producing $\neffsamples=10$ samples, it may appear that \cref{eqn:timing} would predict a $\sim 3$~minute analysis time.
However, the burn-in inefficiency would be increased leading to a decrease in the efficiency and hence increase in the overall run time.
This may still be preferable, but we urge users to give consideration to the efficiency before starting analyses.
We comment that the inefficiencies commented on above are specific to the naive approach of combining multiple independent runs.
The fundamental issue arises that a standard MCMC chain produces unbiased samples from the target density as the number of steps tends to infinity.
There are sophisticated approaches which will overcome these inefficiencies by instead aiming to obtain unbiased samples as the number of MCMC chains tends to infinity \citep{jacob_2020}.
These approaches could be used in future to improve the efficiency of \bilbyMCMC when parallelised over many cores.

In this section, we have seen that \bilbyMCMC can be parallelised both by combining independent runs and utilising multiprocessing.
By comparison, the run-time of the \dynesty nested sampler can only be reduced by the use of multiprocessing.
This is because it is not possible to configure a nested sampler to run part of the full analysis (i.e. only to produce a small subset of the required total number of independent samples).
In this work, we utilize multiprocessing of the \dynesty sampler in \cref{sec:gw} which can take advantage of multi-core processors.
We note that \dynesty can also be used in High Performance Computing (HPC) environments by multiprocessing using many multi-core processors \citep{pbilby}.
We discuss the relative merits of these two approaches with reference to a specific example in \cref{sec:BBHA}.

\section{Standard validation tests}
\label{sec:validation}

In this section, we outline a suite of tests designed to validate the \bilbyMCMC package for standardised problems.
These tests build on previous validation tests of gravitational-wave samplers \citep{Veitch2015, Biwer2019} and tests of the \dynesty sampler \citep{dynesty} implemented in \bilby \citep{Bilby1, Bilby2}.
Though not reported here, we additionally perform integration checks on individual aspects of the sampler and verify that when the likelihood is uninformative the prior is properly recovered.
The scripts used to perform all verification checks and additional figures are available from \href{https://git.ligo.org/gregory.ashton/bilby_mcmc_validation}{git.ligo.org/gregory.ashton/bilby\_mcmc\_validation}; in \cref{tab:all}, we also link to the individual tests.

\begin{table*}
    \centering
    \begin{tabular}{l|lll|rrrrrr}
         Test & Sampler & Configuration & \ntemps & Evidence & JSD [mb] & ACT ($\tau$) & Efficiency & $\nlikelihood/10^6$ & $\neffsamples$ \\ \hline\hline
         \multirow{4}{1.5cm}{\href{https://git.ligo.org/gregory.ashton/bilby_mcmc_validation/-/tree/master/Standard_Normal}{Standard Normal}} & \dynesty & $\nlive{=}2000,\nact=50$ & --- & $0.02\pm0.04$ & 0.9 & --- & 0.58\% & 1.1 & 6200\\
         & \bilbyMCMC & AG-DE-UN & 1 & --- & 0.5 & 6 & 15.0\% & 0.05 & 8000 \\
         & \bilbyMCMC & AG-DE-UN & 16 & $0.03\pm 0.01$ & 0.5 & 5 & 1.2\% & 0.5 & 6000 \\
         & \bilbyMCMC & AG-DE-UN & 32 & $-0.007\pm0.01$ & 0.5 & 5 & 0.6\% & 0.8 & 5000 \\
         \hline
         \multirow{8}{1.5cm}{\href{https://git.ligo.org/gregory.ashton/bilby_mcmc_validation/-/tree/master/Rosenbrock}{Rosenbrock}} & \dynesty & $\nlive{=}2000,\nact=10$ & --- & $0.08\pm0.07$ & 2.7 & --- & 0.7\% & $1.1$ & 7500 \\
         & \dynesty & $\nlive{=}2000,\nact=50$ & --- & $0.04\pm0.07$ & 1.6 & --- & 0.1\% & $6.7$ & 7500 \\
         & \bilbyMCMC & AG-DE-UN-GM-NF-KD & 16 & $-0.02 \pm 0.01$ & 0.7 & 10 & 0.6\% & 3.3 & 20000 \\
         & \bilbyMCMC & AG-DE-UN-GM-NF-KD & 1 & --- & 0.3 & 16 & 6.2\% & 0.34 & 20000 \\
         & \bilbyMCMC & AG-DE-UN-NF & 1 & --- & 0.3 & 19 & 5.2\% & 0.40 & 21000 \\  
         & \bilbyMCMC & AG-DE-UN-KD & 1 & --- & 0.3 & 110 & 0.9\% & 2.2 & 20000 \\  
         & \bilbyMCMC & AG-DE-UN-GM & 1 & --- & 0.3 & 17 & 6.1\% & 0.37 & 22000 \\  
         & \bilbyMCMC & AG-DE-UN & 1 & --- & 0.5 & 171 & 0.6\% & 3.4 & 20000 \\  
         \hline
         \multirow{5}{1.5cm}{\href{https://git.ligo.org/gregory.ashton/bilby_mcmc_validation/-/tree/master/Unimodal_15D_Gaussian}{Unimodal Gaussian}} & \dynesty & $\nlive{=}2000,\nact=10$ & & $-0.05\pm0.2$ & 0.8 & --- & 0.09\% & 22 & 20000\\
         & \dynesty & $\nlive{=}2000,\nact=50$ & --- & $0.07\pm0.2$ & 0.7 & -- & 0.01\% & 150 & 20000 \\
         & \bilbyMCMC & AG-DE-UN-GM-NF-KD & 1 & --- & 0.05 & 85 & 1.2\% & 0.45 & 5000 \\
         & \bilbyMCMC & AG-DE-UN-GM-NF-KD & 16 & $-0.25\pm0.13$ & 0.006 & 70 & 0.09\% & 5.7 & 5000 \\
         & \bilbyMCMC & AG-DE-UN-GM-NF-KD & 32 & $-0.03\pm0.06$ & 0.003 & 61 & 0.05\% & 10 & 5000\\
         \hline
         \multirow{2}{1.5cm}{\href{https://git.ligo.org/gregory.ashton/bilby_mcmc_validation/-/tree/master/Bimodal_15D_Gaussian}{Bimodal Gaussian}} & \dynesty & $\nlive{=}2000,\nact=50$ & --- & $0.02\pm0.2$ & 0.4 & --- & 0.005\% & 390 & 20000 \\
         & \bilbyMCMC & AG-DE-UN-GM-KD & 16 & $-0.05\pm 0.1$ & 0.3 & 3 & 0.02\% & 32 & 5000 \\
         \hline
         \rowcolor{Gray}
         & \dynesty & $\nlive{=}2000,\nact=50$ & --- & $51.4\pm0.2$ & --- & --- & 0.010\% & 160 & 16000
         \\
         \rowcolor{Gray}
         \multirow{-2}{2cm}{\href{https://git.ligo.org/gregory.ashton/bilby_mcmc_validation/-/tree/master/BBH_A}{BBH A}} & \bilbyMCMC & --- & 8 & $49.7\pm 0.2$ & --- & $7\times10^{3}$ & 0.0018\% & 300 & 5000
         \\
         \hline
         \rowcolor{Gray}
         & \dynesty & $\nlive{=}2000,\nact=50$ & --- & $133.4\pm0.2$ & --- & --- & 0.00686\% & $220$ & 15000
         \\
         \rowcolor{Gray}
         \multirow{-2}{2cm}{\href{https://git.ligo.org/gregory.ashton/bilby_mcmc_validation/-/tree/master/BNS_A}{BNS A}} & \bilbyMCMC & --- & 8 & $132.6\pm 0.3$ & --- & $60\times10^3$ & 0.00021\% & 2500 & 5000 \\
         \hline
    \end{tabular}
    \caption{Validation tests reported in this work.
    White-shaded rows are those from the standard validation tests (\cref{sec:validation}) while gray-shaded rows are tests from gravitational wave validation tests (\cref{sec:gw}).
    For the standard validation tests we give the \bilbyMCMC configuration by the set of proposals (described in \cref{sec:proposals}) while for the \dynesty sampler we give $\nlive$ and $\nact$ (cf \citet{Bilby2}).
    For the gravitational-wave validation tests, we use the proposal set described in \cref{tab:gw-proposal-set}.
    In the Evidence column, we report $\Delta \lnZ=\lnZ - \lnZ'$ for the standard validation tests where the exact evidence $\lnZ'$ is known; for the gravitational-wave validation tests (gray rows), where the evidence is not known, we report the natural logarithm of the signal vs. Gaussian noise Bayes factor.
    Where the posterior distribution can be directly sampled from, we report the maximum JSD (see \cref{sec:js}) in milli-bits [mb]. 
    For the MCMC configurations, we list the final-estimated ACT $\tau$; this is always given in raw steps (i.e. we rescale runs which use $\Lone > 1$).
    For the gravitational-wave validation tests, $\tau$ is given by the mean valued averaged over all independent runs, typically this varies by several tens of percent.
    We also report the posterior sampling efficiency described in \cref{sec:efficiency}, the total number of likelihood evaluations, and the number of independent samples the analysis produced.
    }
    \label{tab:all}
\end{table*}

\subsection{Standard Normal distribution}
\label{sec:standard-normal}
As an initial test, we evaluate a one-dimensional standard-normal likelihood where
\begin{align}
    \likelihood(\theta) = \frac{e^{-\theta^2/2}}{\sqrt{2\pi}}\,,
\end{align}
and the prior is uniform between -10 and 10:
\begin{align}
    \pi(\theta) = U(-10, 10)\,.
\end{align}
In this case, the evidence can be estimated as:
\begin{align}
    \Z = \int_{-\infty}^{\infty} \likelihood(\theta)\pi(\theta)\, d\theta \approx \frac{1}{20}\,.
    \label{eqn:sn_evidence}
\end{align}
and the posterior $p(\theta)$ is a standard-normal distribution.

Running the \bilbyMCMC and \dynesty samplers on this problem, in \cref{tab:all}, we report the configurations, the difference in log-evidence, and quantities related to the performance.

To verify the posterior sampling, we calculate the Jensen-Shannon divergence (JSD, see \cref{sec:js} for an extended discussion) between 5000 independent posterior samples drawn using the sampler and samples drawn directly from the known posterior.
For all configurations, we report JSD values below a threshold of \SI{2}{mb} (where mb is the shorthand for a milli-bit of information): this demonstrates the posteriors are \emph{statistically identical}.
As such, we conclude that both the \dynesty and \bilbyMCMC samplers are able to sample this simple inference problem without bias and report accurate estimates of the uncertainty.

To verify the estimates of the Bayesian evidence, we compare with the known evidence calculated in \cref{eqn:sn_evidence}.
Both the \dynesty and \bilbyMCMC sampler using $\ntemps=32$ produce estimates of the evidence that agree with \cref{eqn:sn_evidence} to within the stated uncertainties.
However, it is known that parallel-tempered evidence estimates have a bias which can be reduced by increasing the number of temperatures \citep{Xie2010, russel2019}.
This point is demonstrated by the \bilbyMCMC analyses with \ntemps=16 which does not produce a result consistent with the known evidence.

\subsection{Rosenbrock likelihood}
\label{sec:rosenbrock}
We analyze the Rosenbrock likelihood~\citep{Rosenbrock1960}, taking the explicit form and priors from Eqn.~(C2) of \citet{Fowlie2020}.
The banana-shaped posterior is challenging to sample from and representative of the types of posteriors seen in CBC inference problems.
This makes it an ideal validation test.
Results for several configurations of both samples are listed in \cref{tab:all}.

We sample directly from the posterior distribution of the Rosenbrock likelihood using a reparameterization. This enables us to calculate the maximum JSD between samples drawn using different configurations of the \bilbyMCMC and \dynesty samplers and the directly-sampled posterior.
The maximum JSD for the \bilbyMCMC analyses all fall below the \SI{2}{mb} threshold for statistically identical posteriors.
However, we find that the samples from the \dynesty sampler using $\nact=10$ are marginally above this threshold while the analysis with $\nact=50$ is below.
In re-running these analyses, we find variations in the JSD value of order $\sim50\%$: this indicates the \dynesty analyses are subtly biased.
\nact is a user-controlled parameter described in \citet{Bilby2} which determines the number of internal MCMC steps to take based on the estimated autocorrelation time.
A value of 10 was previously found to be sufficient for binary black hole analyses \citep{Bilby2}, but this test demonstrates larger values may be necessary to ensure convergence for the Rosenbrock likelihood.
The dependence on \nact indicates the cause is likely to be the MCMC-within-nested-sampling algorithm itself (we used the version in Bilby v1.1.3 for the analyses in this work); investigation is needed to determine if this is failing and to resolve this bias. 

\begin{figure}
    \centering
    \includegraphics[width=\columnwidth]{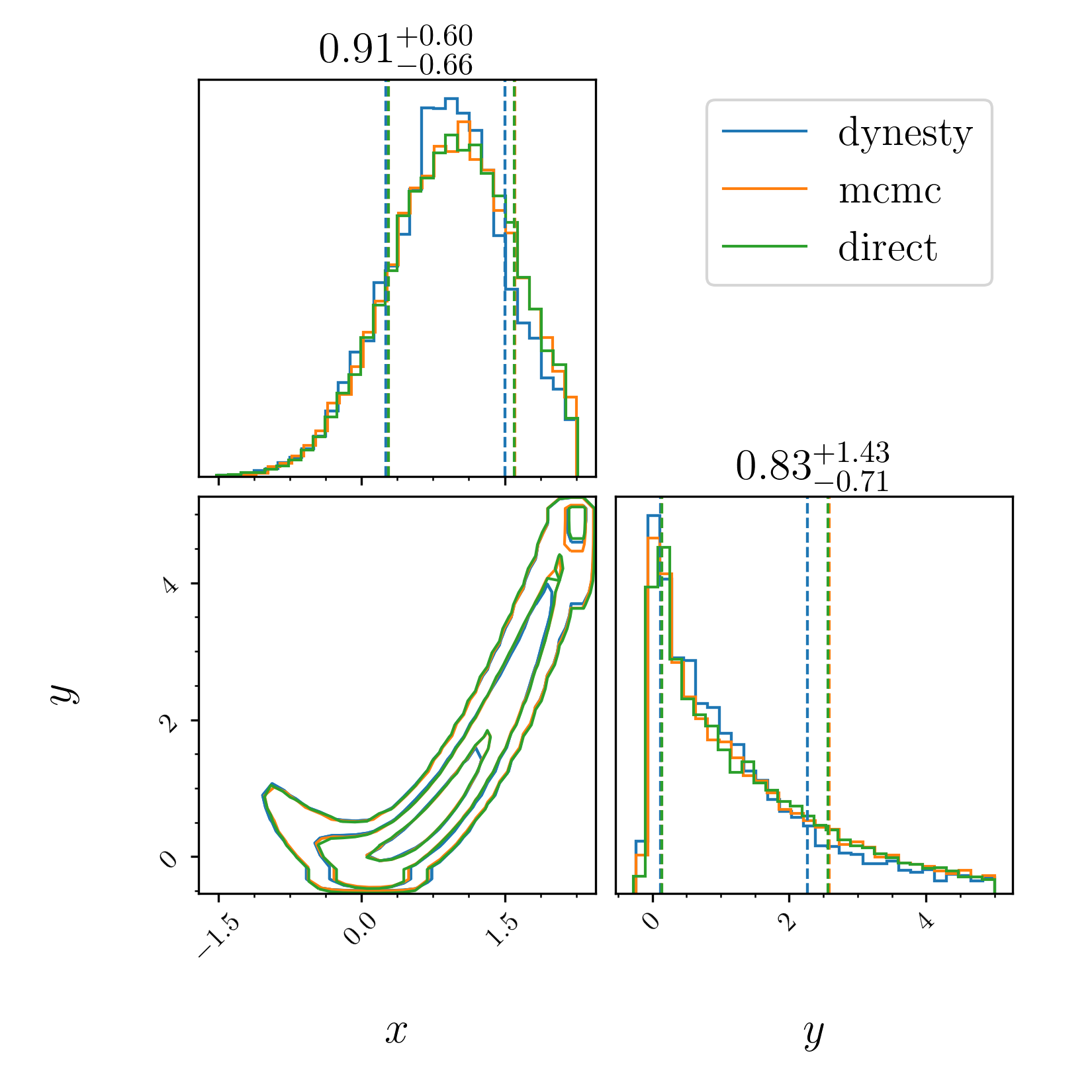}
    \caption{Comparison of the \dynesty sampler (with $\nlive=2000$ and $\nact=50$), the \bilbyMCMC sampler (with $\ntemps=1$ and the full set of learning proposals), and samples drawn directly from the posterior for the Rosenbrock test.
    The JSD test between each of the samplers and the direct samples (see \cref{tab:all}) quantifies that the \bilbyMCMC sampler produces statistically identical posterior samples while the \dynesty sampler produces JSD values at the failure threshold.
    Visually, we see that the posterior samples produced by the \dynesty sampler are overly constrained.
    }
    \label{fig:rosenbrock}
\end{figure}

We visualise the results in Fig.~\ref{fig:rosenbrock}: \bilbyMCMC and the ``direct'' samples agree, but samples from the \dynesty analyses (with $\nact=50$) are overly constrained.
This is a typical failure mode of posterior samples generated by nested sampling methods which use bounding ellipsoids to improve performance of the sampler.
We note that we do not see similar issues for CBC inference problems (see \cref{sec:BBHA} and \cref{sec:BNSA}).
This failure requires further investigation and highlights the need for cross-sampler comparisons.

The various MCMC configurations in \cref{tab:all} enable a comparison of the impact of the learning proposals.
Using all three learning proposals (AG-DE-UN-GM-NF-KD), reduces the ACT by a factor of $\gtrsim 10$ with respect to the analysis without any learning proposals (AG-DE-UN).
By running each of the learning proposals individually, we see that the normalizing flows and GMM proposals both have a similar performance improvement (with respect to the AG-DE-UN proposals alone) to all three together.
Meanwhile the KDE proposal alone provides only a factor of $\sim2$ reduction in the ACT.
This demonstrates that learning-proposals are a powerful tool in improving the efficiency of the MCMC algorithm.

Finally, we turn to evidence estimation. The Rosenbrock likelihood used in this work has an analytically approximated evidence of $\lnZ'=-5.804$~\citep{Fowlie2020}.
In \cref{tab:all}, we provide evidence estimates for the \dynesty and \bilbyMCMC sampler with $\ntemps=16$.
For the \dynesty sampler, the evidences agree to within the stated uncertainties.
For the \bilbyMCMC sampler, the evidence estimate disagrees at the level of 1 standard deviation.
This performance is consistent with the findings of \citet{Veitch2015} in which the LALInference MCMC sampler similarly struggled to consistently estimate the evidence of the Rosenbrock likelihood.

\subsection{15-dimensional unimodal Gaussian}
\label{sec:unimodal}
\begin{figure}
    \centering
    \includegraphics{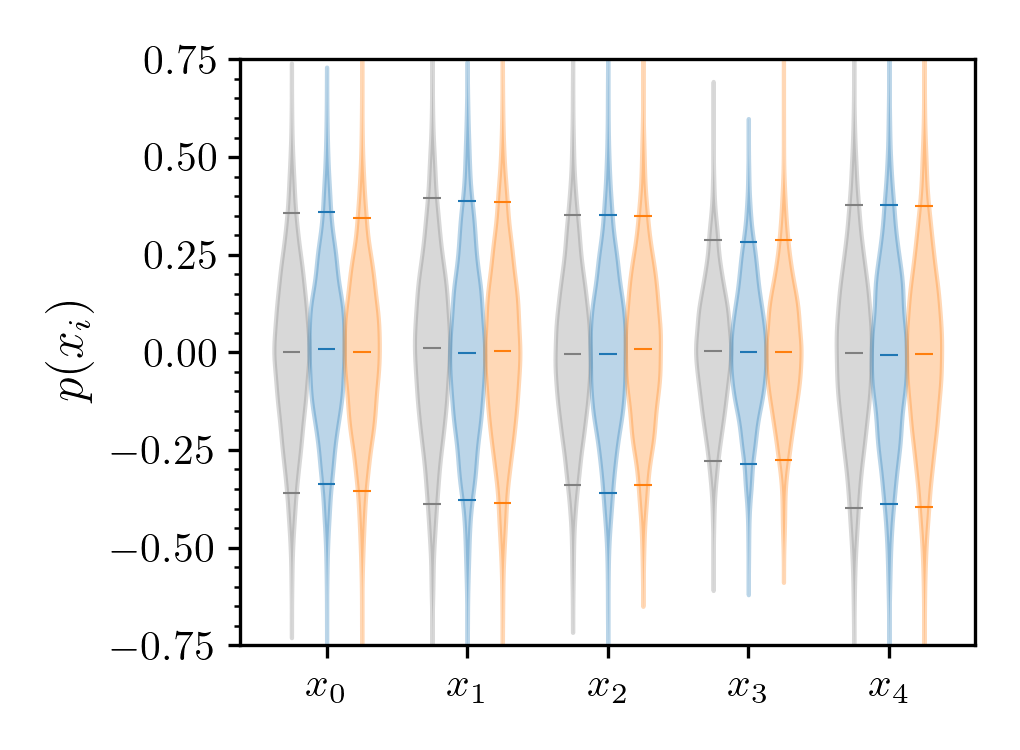}
    \caption{Violin plot showing posteriors from 5 parameters of the 15-dimensional unimodal Gaussian test. Each violin shows 5000 samples drawn directly from the posterior (gray), from the \dynesty analysis with $\nact=50$ (blue) and the \bilbyMCMC analysis with $\ntemps=1$ (orange). Vertical lines denote the median and 90\% credible interval.}
    \label{fig:violin}
\end{figure}

We analyze the 15-dimensional unimodal multivariate Gaussian distribution originally proposed in \citet{Veitch2015} using the specific configuration from \citet{Bilby2}.
We report the results in \cref{tab:all}, varying the number of parallel-tempered chains, but utilising the standard proposal sets.
For all samplers and configurations, the maximum JSD falls below the nominal 2~mb threshold for statistically identical samples.
To visualise the results, in \cref{fig:violin} we plot a subset of 5 posteriors in a violin plot.
This shows strong agreement between the samplers and with samples drawn directly from the posterior.

The evidence for this 15-D unimodal Gaussian test case can be estimated directly \citep{Bilby2} as $\lnZ' \approx-34.54$.
The \dynesty sampler correctly estimates the evidence to within the stated uncertainty in both configurations.
Meanwhile, the \bilbyMCMC sampler gets close to the true evidence, with $\ntemps=32$, but suffers the previously-discussed bias when $\ntemps$ is small.

We study the performance of \bilbyMCMC while varying $\ntemps$.
In this unimodal case, even a single-temperature sampler can sample the posterior.
Increasing the number of parallel temperatures from 1 to 16 marginally reduces the ACT.
As \ntemps is increased, the uncertainty on the evidence estimate decreases, but the ACT does not significantly change.
This is expected since it is a unimodal target density which does not require parallel-tempering to hop between modes.
As such, increasing the number of temperatures (to achieve a meaningful estimate of the evidence) results in a reduction in the efficiency as predicted by \cref{eqn:burn-in-inefficiency}.

\subsection{15-dimensional bimodal Gaussian}
\label{sec:bimodal}
We analyse a bimodal Gaussian distribution consisting of two copies of the unimodal Gaussian distribution (cf. \cref{sec:unimodal}) and means separated by 8 standard deviations in each dimension (as used in \citet{Bilby2}).
This test probes the ability of the sampler to efficiently hop between modes. 
With a single cold chain, the MCMC sampler is unable to find both modes (in other words, the ACT is infinite).
With $\ntemps=16$, \bilbyMCMC is able to sample from both modes.
Comparing to samples drawn directly from the posterior, the maximum JSD for the \dynesty sampler and \bilbyMCMC sampler both fall below the threshold for statistically identical samples.
\citet{Bilby2} noted that the \dynesty sampler tends to overweight one or other of the two modes in this test.
But, that combining over many runs the effect averages out.
We confirm this in our individual run of the \dynesty sampler.
For the \bilbyMCMC sampler, we find that the effect is weaker.
Quantifying the effect by the number of samples in each mode, the \bilbyMCMC tends to produce more equal-weighted posteriors (in agreement with the true posterior).
This can be understood because the MCMC sampler is proposing jumps between modes while for the \dynesty sampler the relative weights of the two modes is determined by the bounding ellipsoids.

The evidence for the 15-D bimodal Gaussian be directly estimated \citep{Bilby2} as $\lnZ'\approx-34.54$.
Comparing the evidence estimated by the samplers to this direct estimation, we find similar performance to that of the 15-D unimodal Gaussian studied in \cref{sec:unimodal}.
Namely, the \dynesty sampler outperforms \bilbyMCMC in accuracy and uncertainty.

\section{Gravitational-wave validation tests}
\label{sec:gw}

In this section, we discuss the specifics and validation of the \bilbyMCMC sampler for CBC gravitational-wave inference.
The inference of CBC coalescence signals has been well-studied in the literature. The fundamentals can be found in \cite{Veitch2015}, a recent review in \citet{thrane2019}, and the specifics of the \bilby interface in \citet{Bilby1} and \citet{Bilby2}.
In \cref{sec:parameterisation}, we introduce the basics of the CBC model and describe the best-known parameterisation of $\theta$ to reduce the ACT.
Then, in \cref{sec:marginalization}, we discuss the use of analytic marginalization methods which reduce the dimensionality of $\theta$ in sampling.

\subsection{Models, optimal parameterisation, and priors}
\label{sec:parameterisation}
A circularised gravitational-wave signal from a CBC can be described by a set of 17 model parameters $\theta$.
We can partition $\theta$ into 11 intrinsic parameters (two mass, six spin parameters, the binary phase, and up to 2 tidal deformability parameters) and 6 extrinsic parameters (the 3-D localisation, polarisation, merger time, and the angle between the total angular momentum and the line of sight).

There are many ways to choose these 17 parameters in the literature.
These different parameterisations offer varying levels of computational convenience and interpretability.
In this work we use the following parameterisation for CBC analyses based on which parameters lead to the shortest auto-correlation lengths in our tests.

{\em Mass.}
Labelling the detector-frame mass of the two objects in the binary $m_1$ and $m_2$, we sample in the detector-frame chirp mass
\begin{align}
    \mchirp = \frac{(m_1m_2)^{3/5}}{(m_1 + m_2)^{1/5}}\,
\end{align}
and mass ratio $q = \frac{m_2}{m_1}$.
We apply prior cuts, discussed below, such that $q \le 1$.
This is the standard choice employed for compact binary analyses as the chirp mass is the best measured parameter for binary inspirals, followed by the mass ratio \citep{cutler1994}.

{\em Spin.}
The spin of the compact objects contribute six degrees of freedom to the problem.
Following \citep{2014PhRvD..90b4018F}, we sample in the \emph{magnitudes and tilts} parameterised in spherical coordinates with the $z-$axis aligned with the total angular momentum using the magnitude $a_i$ and tilt $\theta_i$ (where $i \in [1, 2]$ labels the primary and secondary objects) along with two azimuthal parameters $\{\phi_{jl}, \phi_{12}\}$.

{\em Tides.}
Tidal deformability of neutron stars is typically described in terms of either two dimensionless deformability parameters $\Lambda_i$ or combinations of these two combinations of these parameters that directly determine the contribution to the phase contribution $\{\tilde{\Lambda}, \delta \tilde{\Lambda}\}$ \citep{flanagan2008, favata2014, wade2014}.
We sample in the latter set as these reduce the ACT.

{\em Location.}
The location of the binary is uniquely described by four parameters, the distance to the source, the two-dimensional sky location, and the merger time.
We choose these parameters as in \LALInference.
We specify the merger time as the time of arrival of the merger signal at one of the detectors, ideally the one where we expect the highest signal-to-noise ratio.
We characterize the sky location using a reference frame based on the separation vector of two of the detectors;
see \cite{Bilby2} Sec.~3.1.7 for an explicit definition.
For all of the results presented here, we marginalize over the distance to the source (see \cref{sec:marginalization}), so the specific choice of distance parameter is irrelevant.

{\em Orientation.}
Finally, we require three Euler angles to convert the binary frame to the galactic reference frame.
These are the inclination angle between the binary angular momentum and the line-of-sight from the source to the observer $\thetajn$, the binary phase at a reference frequency $\phi$ (typically the merger frequency), and the polarization of the source $\psi$.
Throughout, we sample in $\cos(\thetajn)$ and $\psi$.
In practice there is a strong correlation between the phase and the polarization and so we sample in a phase offset parameter:
\begin{equation}
    \delta \phi = \begin{cases}
    \phi + \psi & \thetajn \leq \frac{\pi}{2}
    \\
    \phi - \psi & \thetajn > \frac{\pi}{2}
    \end{cases}.
    \label{eqn:delta_phase}
\end{equation}
The change of sign is due to a change in the direction of the degeneracy when observing from above/below the orbital plane.
This parameterisation introduces a discontinuity in the likelihood at $\theta_{JN} = \pi / 2$. However, the proposal schemes outlined in \cref{sec:proposals}, including the machine-learned proposals, do not depend on assumptions of smoothness.
In practice we find that the parameterisation improves the performance relative to analyses which use the phase directly.

Following \LALInference, we apply a prior uniform on the component masses $m_1$ and $m_2$ with cuts in the chirp mass and mass ratio.
We then apply the non-informative priors on all other parameters and a uniform in the source-frame prior for the luminosity distance \citep{Bilby2}.

\subsection{Analytic likelihood marginalization}
\label{sec:marginalization}

Of the 17-dimensional parameter space described in \cref{sec:parameterisation}, there are three, the luminosity distance, geocentric time, and binary phase over which we are able to efficiently marginalize the gravitational-wave likelihood [see \citet{veitch13, farr14, Veitch2015, singer16a, singer16b} and \citet{thrane2019} for a review].
In the context of an MCMC sampler, the marginalized likelihood has a shorter ACT relative to the non-marginalized likelihood.
This is both due to the reduction in dimensionality and to the reduction in the complexity of the posterior.
Since it is possible to reconstruct the marginalized parameters after analysis \citep{thrane2019}, where possible marginalized likelihoods are strongly recommended.
For the luminosity distance, we always marginalize the likelihood.
For the geocentric time, we marginalize the likelihood [and add the time jitter, $t_{j}$ as described in \citet{Bilby2}] except in instances where the reduced-order-quadrature method ROQ is used in which time-marginalization has not yet been implemented. 
The assumptions made in marginalizing the binary phase are invalid for precessing CBC systems or models that include higher-order emission modes.
Therefore, we do not marginalize the binary phase in this work.
But, in future use cases, where a non-pressing waveform without higher-order emission modes is considered, we do recommend using phase marginalization.

\subsection{Fiducial binary black hole: BBH A}
\label{sec:BBHA}

We simulate a fiducial (reference) binary black hole (BBH) signal observed by the LIGO Hanford and Livingston detectors \citep{2015CQGra..32g4001L} at their design sensitivity \citep{abbott_2020_obs_scen}.
The simulation parameters, labelled as BBH A, are given in \cref{tab:simulations}.
We use the \texttt{IMRPhenomPv2} \citep{hannam2014, schmidt2012} waveform approximant to both simulate and analyse the signal.
In this noise realisation, the simulated signal has a network matched-filter signal to noise ratio (SNR) of $\sim13$. 

\begin{table}
    \centering
    \begin{tabular}{|l|l|l|c|c|}
         \hline
         \multicolumn{3}{|c|}{Parameters} & BBH A & BNS A \\ \hline\hline
         & & \mchirp &  17.1 & 1.4875\\
         & \vertlabel{-2}{Mass} & \massratio &  0.62 & 0.950\\
         \cline{2-5}
         & & $a_1$ &  0.296 & 0.01\\
         & & $a_2$ &  0.393 & 0.01\\
         & & $\theta_1$ &  0.09 & 0\\
         & & $\theta_2$ &  1.20 & 0\\
         & & $\phi_{12}$ &  1.10 & 0\\
         & \vertlabel{-6}{Spin} & $\phi_{jl}$ & 0.52 & 0\\
         \cline{2-5}
         & & $\Lambda_1$ & 0 & 1500\\
         \vertlabel{-11}{Intrinsic} & \vertlabel{-2}{Tidal} & $\Lambda_2$ &  0 & 750\\
         \hline
         & & RA &  3.95 & 1.67\\
         & & DEC & 0.22 & -1.22\\
         & \vertlabel{-3} {Loc.} & $d_{L}$ &  497 & 180\\
         \cline{2-5}
         & & \thetajn &  1.88 & -0.88\\
         & & $\psi$ &  2.70 & 2.70\\
         \multirow{-6}{*}{\rotatebox[origin=c]{90}{Extrinsic}} & \vertlabel{-3}{Orient.} & $\phi$ &  3.69 & 3.69\\
         \hline
    \end{tabular}
    \caption{Simulation parameters for the three fiducial events analysed in \cref{sec:gw}.
    In the left-two columns, we provide the parameter groups names as described in \cref{sec:parameterisation}.
    }
    \label{tab:simulations}
\end{table}

We analyse \SI{4}{s} of simulated data with the \dynesty and \bilbyMCMC samplers using the configurations described in \cref{tab:all}, the priors described in \cref{sec:gw}, and distance and time marginalization.
For the \bilbyMCMC sampler, we use 13 independent chains, a thinning factor of $\gamma=0.2$, and run each chain until it produces 2000 samples. In total, this produces 25000 samples with $\neffsamples=5000$.
For the \dynesty sampler, we use the standardised configuration listed in \cref{tab:all}, but use two independent run to enable a robustness check.

\begin{figure}
\centering
\subfigure{\includegraphics[width=.22\textwidth]{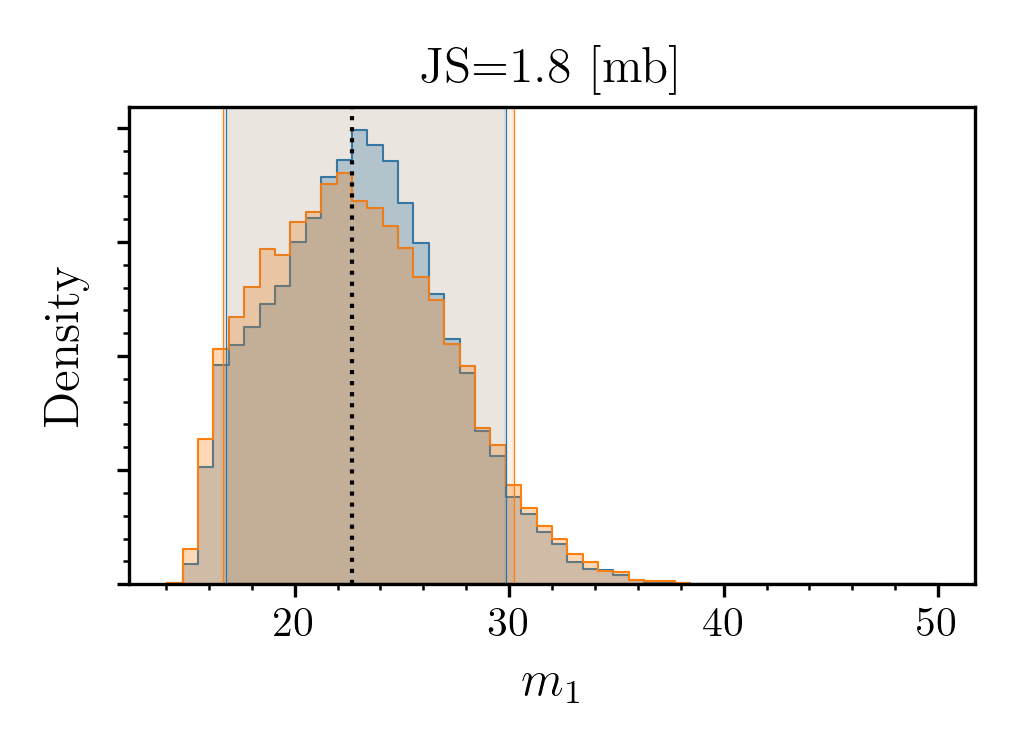}}
\subfigure{\includegraphics[width=.22\textwidth]{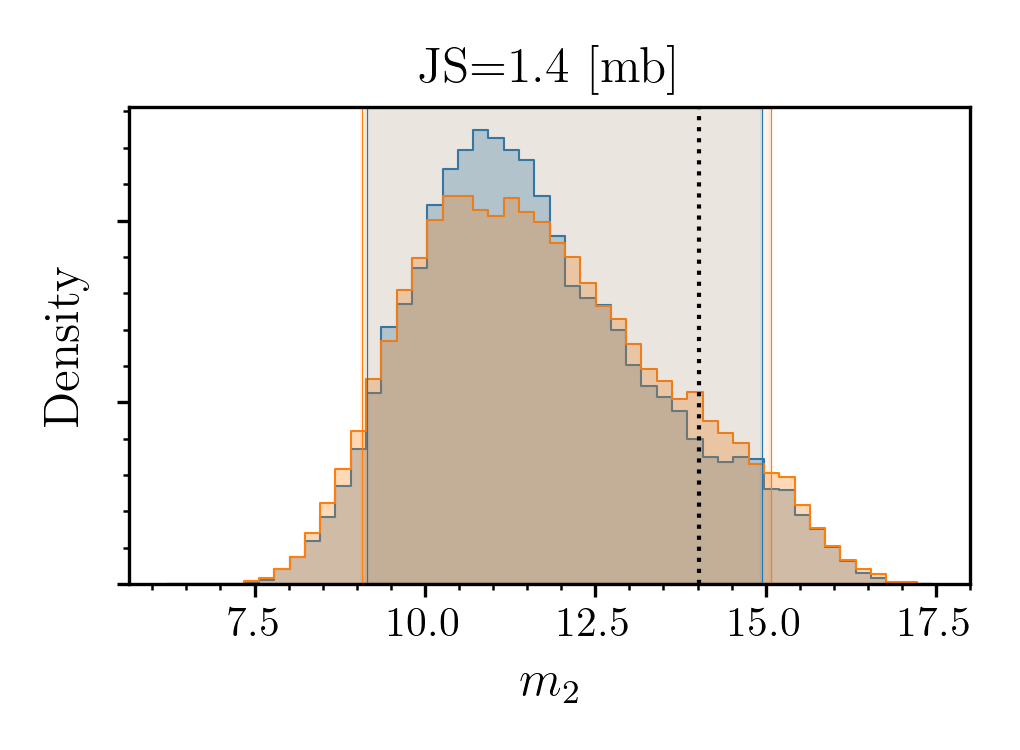}}
\subfigure{\includegraphics[width=.22\textwidth]{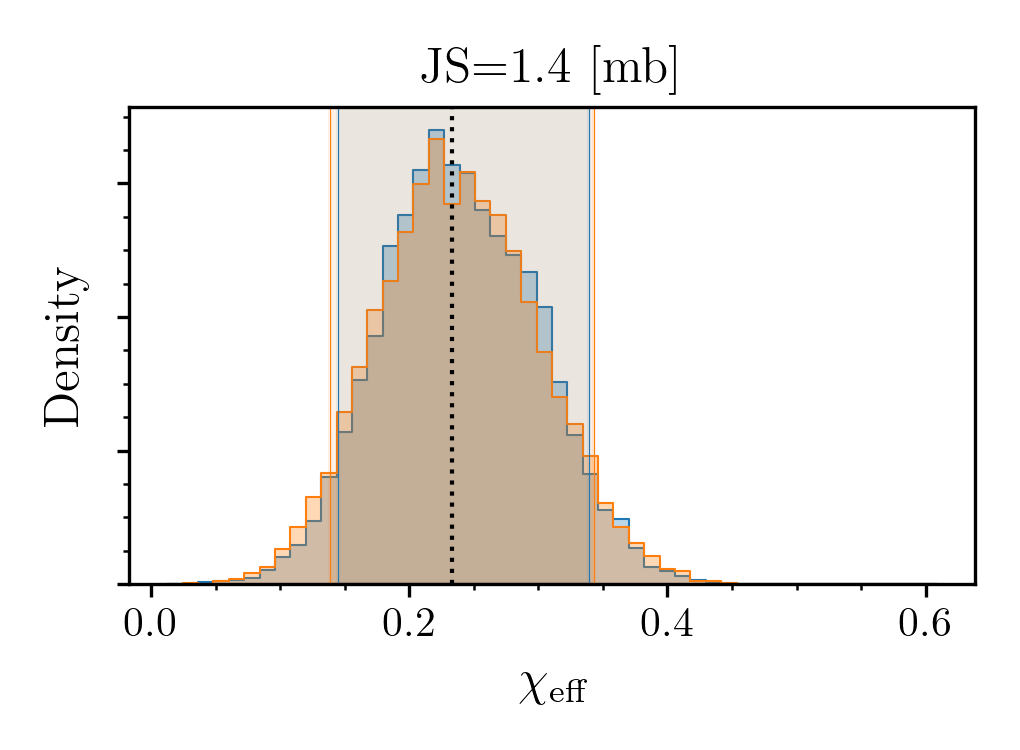}}
\subfigure{\includegraphics[width=.22\textwidth]{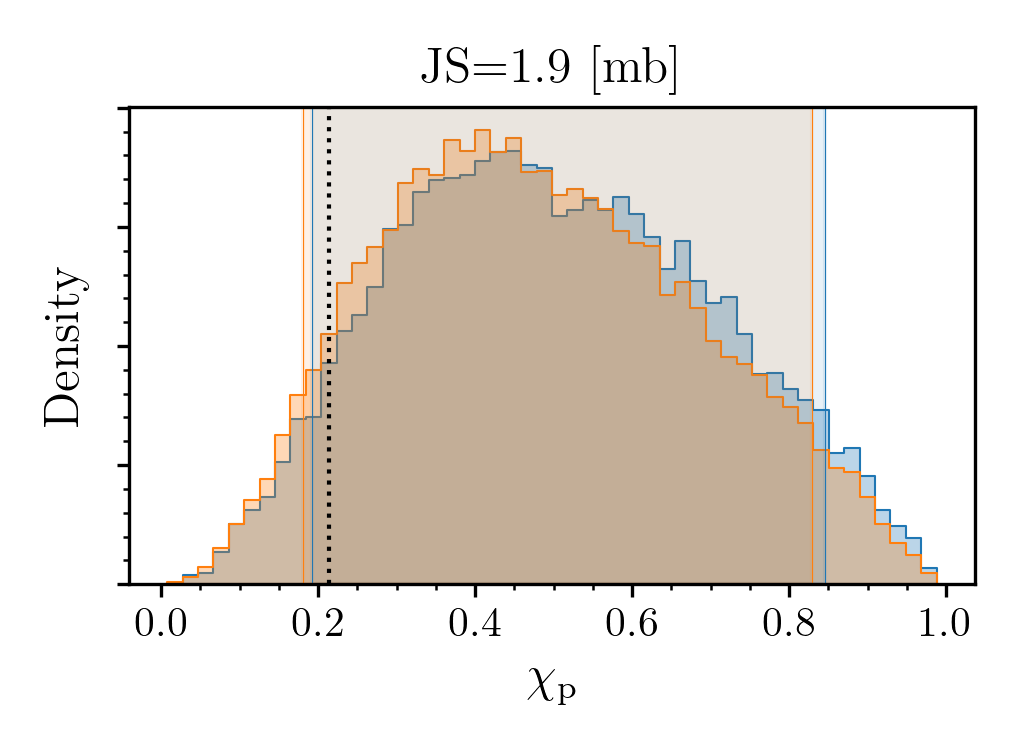}}
\subfigure{\includegraphics[width=.22\textwidth]{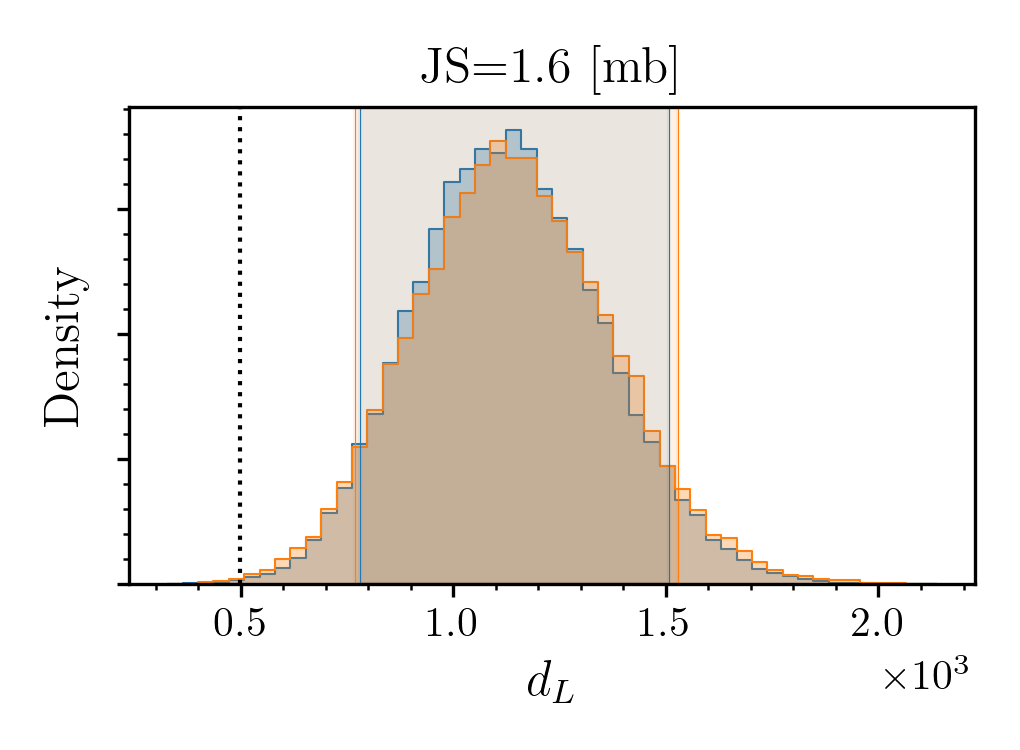}}
\subfigure{\includegraphics[width=.22\textwidth]{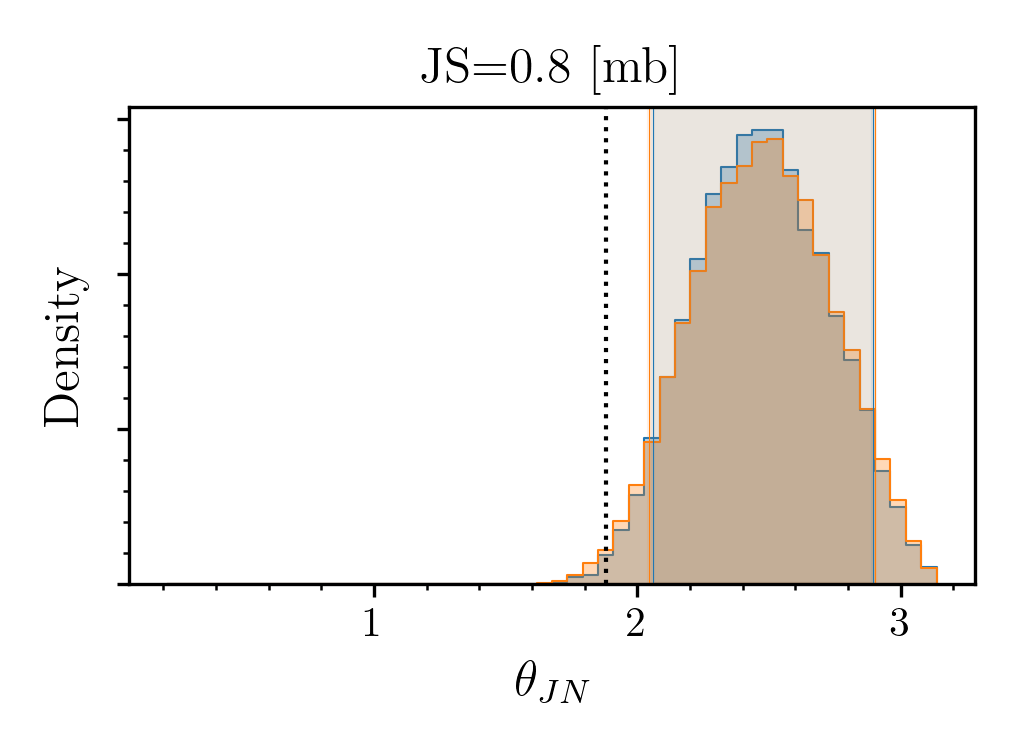}}
\caption{Histograms of the posteriors from the \dynesty (blue) and \bilbyMCMC (orange) analyses of the fiducial BBH. Configurations and summary statistics are given in \cref{tab:all}. Vertical lines mark the edges of the 90\% credible interval for each sampler and a black dotted lines marks the value used to simulate the data.
Note that we do not expect the posteriors to peak at the simulation values due to the influence of the simulated noise and the Bayesian prior.
In the title of each figure, we give the JSD; across all parameters this JSD is found to be below the \SI{2}{mb} threshold we use to determine if the two sets of posteriors are statistically different.
}
\label{fig:bbh}
\end{figure}

It is not possible to sample directly from the posterior in this case, so we resort to cross-sampler comparisons to verify posterior sampling.
Across all CBC parameters, we find that the maximum JSD between the samplers falls below the \SI{2}{mb} threshold: i.e. we find statistically identical posteriors between \dynesty and \bilbyMCMC.
To visualise these difference in \cref{fig:bbh}, we plot histograms of several quantities of astrophysical interest, along with their individual JSD.


In the evidence column of \cref{tab:all}, we report the Bayes factor between the signal evidence and the Gaussian noise evidence (for a fixed realisation of the noise and power spectral density this is a fixed quantity).
The evidence estimates disagree at the 1-$\sigma$ level of the quotes uncertainties.
The difference is likely explained by the known bias in parallel-tempered evidence (cf. \cref{sec:evidence}) when $\ntemps$ is small.
Here, we tune $\ntemps$ for efficient sampling of the posterior, rather than evidence estimation.
In such a configuration, we recommend that the evidence estimate only be used as a rough guide, but not be used for quantitative analysis.
To reduce the bias, \ntemps can be increased at the cost of posterior sampling efficiency.

Comparing the performance of the two samplers, the \dynesty sampler is an order of magnitude more efficient than the \bilbyMCMC sampler.
However, this efficiency does not directly translate into an order of magnitude reduction in wall-time.
To understand why, we need to discuss the parallelisation strategies available.

As discussed in \cref{sec:parallel}, we have two available levels of parallelisation: combining independent runs and multiprocessing using $\ncores$ processors.
For the \dynesty sampler, reductions in wall-time can only be achieved via multiprocessing.
This is because it is not possible to configure a nested sampler to run part of the full analysis (i.e. to only produce a small subset of the required total number of independent samples).
A simple model for the wall-time of the \dynesty run which agrees with our measured wall-time is:
\begin{align}
    T = \SI{28}{hrs}~
    \left(\frac{\nlikelihood}{160\times10^{6}} \right)
    \left(\frac{\tlike}{\SI{10}{ms}} \right)
    \left(\frac{\ncores}{16} \right)^{-1}\,,
    \label{eqn:timing-bbhA-dynesty}
\end{align}
where $\nlikelihood$ is the number of likelihood evaluations (cf. \cref{tab:all}) and $\tlike$ is the approximate per-likelihood evaluation time for the BBH A likelihood.
Here, we use 16-core processors: below we will discuss the potential scaling to larger multiprocessing pools.

On the other hand, for \bilbyMCMC we can parallelise using independent runs and multiprocessing. We run several independent runs, each producing 400 independent samples.
In a HTC environment (and assuming access to resources is not limited), these can be run at the same time so that the total analysis wall time is given by the wall-time of any individual run.
Using \cref{eqn:parallel-timing} and perfectly matching $\ncores$ to $\ntemps$:
\begin{align}
    \ttotal \approx
    \SI{10}{hrs}~
    \left(\frac{\neffsamples}{400}\right)
    \left(\frac{\tlike}{10{\rm ms}}\right)
    \left(\frac{\epsilon}{0.0017\%}\right)^{-1}
    \left(\frac{m}{0.75}\right)^{-1}
    \left(\frac{\ncores}{8}\right)^{-1}\,,
    \label{eqn:timing-bbhA-mcmc}
\end{align}
where we use the actual efficiency from \cref{tab:all} and multi-processing speed-up factor from \cref{sec:parallel}.
Both \cref{eqn:timing-bbhA-dynesty} and \cref{eqn:timing-bbhA-mcmc} agree with the empirically measured values (up to errors expected for varying access to resources in a HTC environment).

The net result is that the \bilbyMCMC sampler is less efficient, but can be setup to enable a shorter wall time by utilising independent runs.
Some of this inefficiency arises from the sampler itself, some from the burn-in inefficiency.
For this configuration, the burn-in inefficiency, \cref{eqn:burn-in-inefficiency}, is a few percent; further parallelisation (in terms of more independent runs) would increase this inefficiency.

For the \dynesty sampler, reducing the wall-time can only be achieved via access to a larger multiprocessing pool.
The ability to do this is restricted by the available hardware: \ncores of 8 to 16 are typical in most HTC environments though modern CPUs with up to 128 cores do exist which could provide significant speed ups.
Beyond this, massively parallelised nested sampling can leverage multiple CPUs in a HPC environment: in \citet{pbilby}, processing pools including several hundred cores have been used providing two orders of magnitude of speed up.
(We caution that we have not verified the validity of \cref{eqn:timing-bbhA-dynesty} for such massively-parallel environments).
However, access to such resources requires synchronised usage of a dedicated HPC environment.

\begin{figure}
    \centering
    \includegraphics[width=\columnwidth]{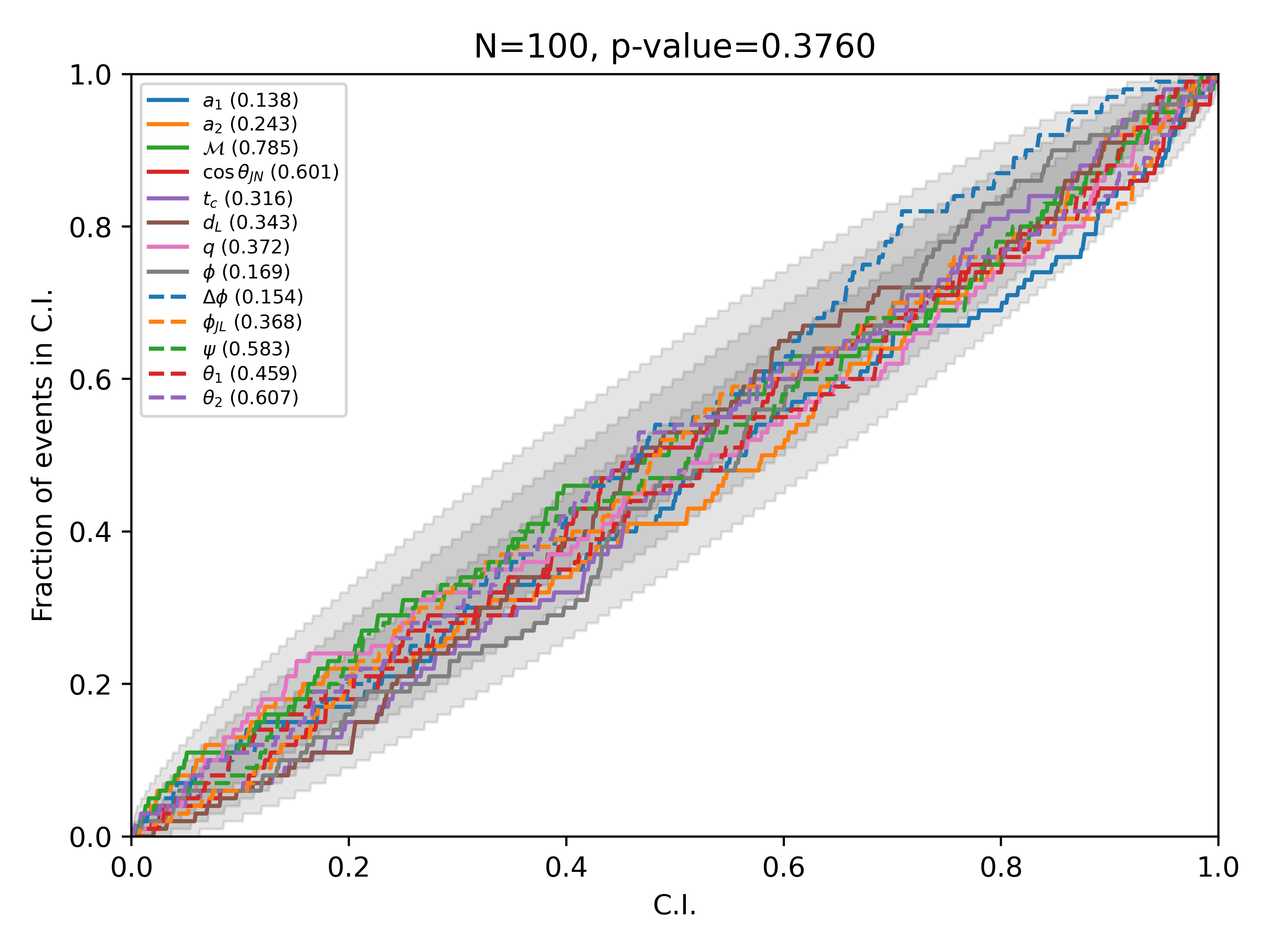}
    \caption{
    A parameter-parameter test for the \bilbyMCMC sampler for simulated BBH systems.
    We plot the fraction of simulated events found within the confidence interval (C.I.) as a function of the C.I.
    For an unbiased sampling from the posterior distribution, lines of this plot are diagonal: we add three gray shaded regions showing the 1, 2, and $3-\sigma$ quantiles.
    To quantify if the results are consistent with an unbiased sampling, we calculate a p-value of the probability that they are unbiased.
    The p-value for each individual parameter is given in the legend and a combined p-value is given in the title.
    Under an unbiased result, we would expect the p-value to be a draw from uniform distribution on [0, 1].
    Since all individual parameters (and the combined result) are greater than $1/15$ (a nominal threshold based on the number of parameters), we conclude the sampler is unbiased, at least at the level probed by 100 simulations.
    }
    \label{fig:pp}
\end{figure}

To investigate the potential for bias in the \bilbyMCMC sampler, in  \cref{fig:pp}, we show the results of a parameter-parameter (PP) test \citep{cook06, talts18} for BBH systems.
This is an important test, typically it fails when one or more of the proposal distributions does not respect detailed balance.
In this test, we simulate 100 BBH signals drawn from an astrophysical prior distribution, analyse each using the \bilbyMCMC sampler, and then check the consistency of the reported credible intervals.
Specifically, \cref{fig:pp} shows the number of events in a given confidence interval as a function of the confidence interval.
We find that the \bilbyMCMC sampler is unbiased at the level probed by this test.

\subsection{Fiducial binary neutron star: BNS A}
\label{sec:BNSA}
We simulate a fiducial binary neutron star (BNS) merger using the \pvtwonrt waveform \citep{dietrich2017, dietrich2019} which includes matter effects from the two neutron stars.
The simulation parameters of the system, BNS A, listed in \cref{tab:simulations} are much lower in mass than that of the BBH systems previously studied.
The result of this lower-mass is that the signal spends a longer duration in the observable band of the detectors (typically, above \SI{20}{Hz}).
To capture this, we analyse \SI{128}{s} of data.
Necessarily, this results in a significant increase in the time required to analyse the likelihood and hence overall wall-time.
To mitigate this, we use the Reduced-Order-Quadrature (ROQ) method
\citep{2012arXiv1210.0577A, 2013PhRvD..87l4005C, 2015PhRvL.114g1104C, smith16, 2020arXiv200913812Q} with the basis provided by \cite{baylor2019} to decrease the per-likelihood evaluation cost.

The simulated signal has small spin components aligned along the angular momentum axis, an arbitrarily selected choice of tidal deformability parameters, and nearly equal mass components. In the specific noise realisation used, the network matched-filter SNR is $\sim18$.
We analyse the signal using both the \dynesty and \bilbyMCMC samplers using the configurations described in \cref{tab:all}. The analyses are identical to those of the BBH A analysis, except, we use the \pvtwonrt waveform model (through the ROQ basis), use only distance marginalization, and restrict the spins to a low-spin configuration (dimensionless spin magnitude less than 0.05 \citep{abbott18_GW170817_properties}).

\begin{figure}
\centering
\subfigure{\includegraphics[width=.22\textwidth]{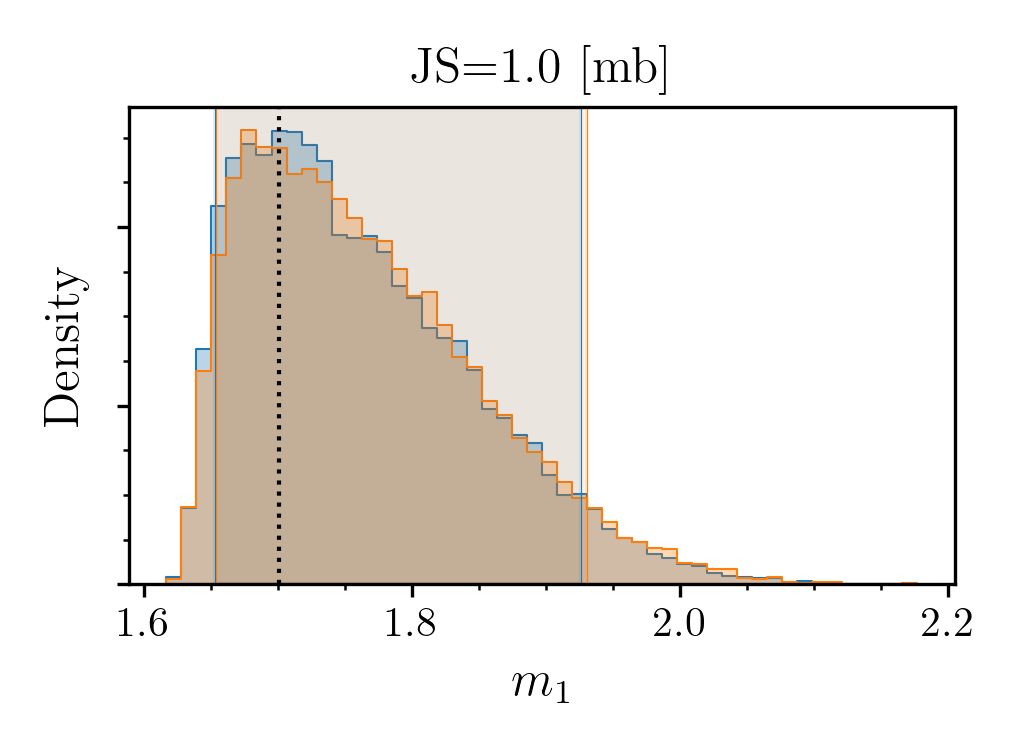}}
\subfigure{\includegraphics[width=.22\textwidth]{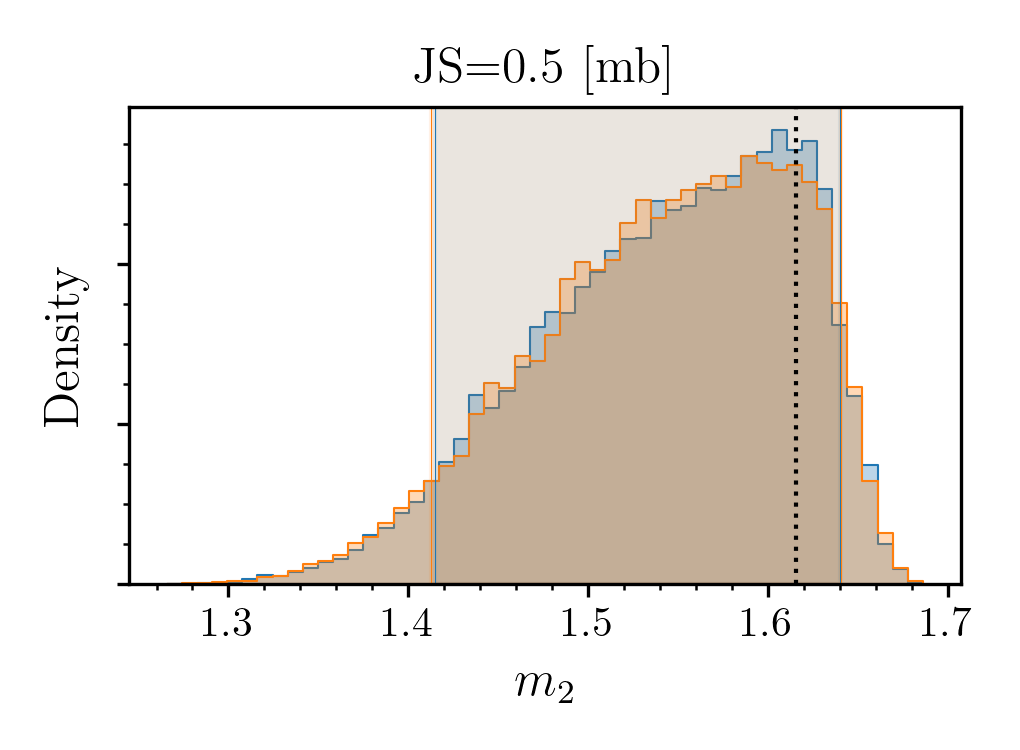}}
\subfigure{\includegraphics[width=.22\textwidth]{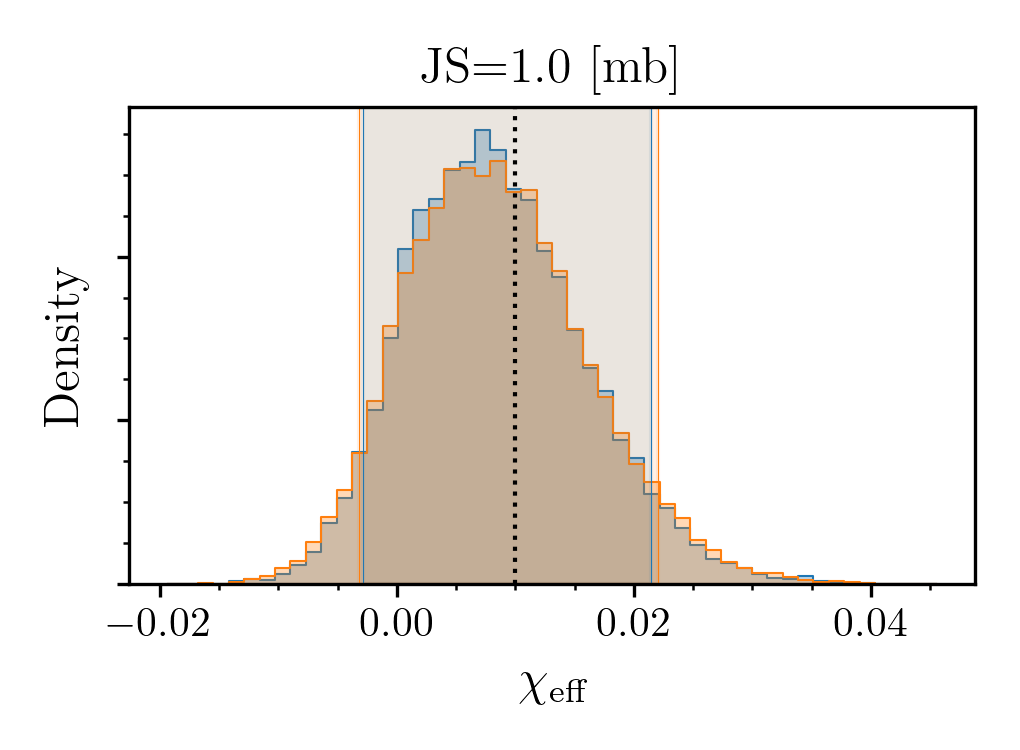}}
\subfigure{\includegraphics[width=.22\textwidth]{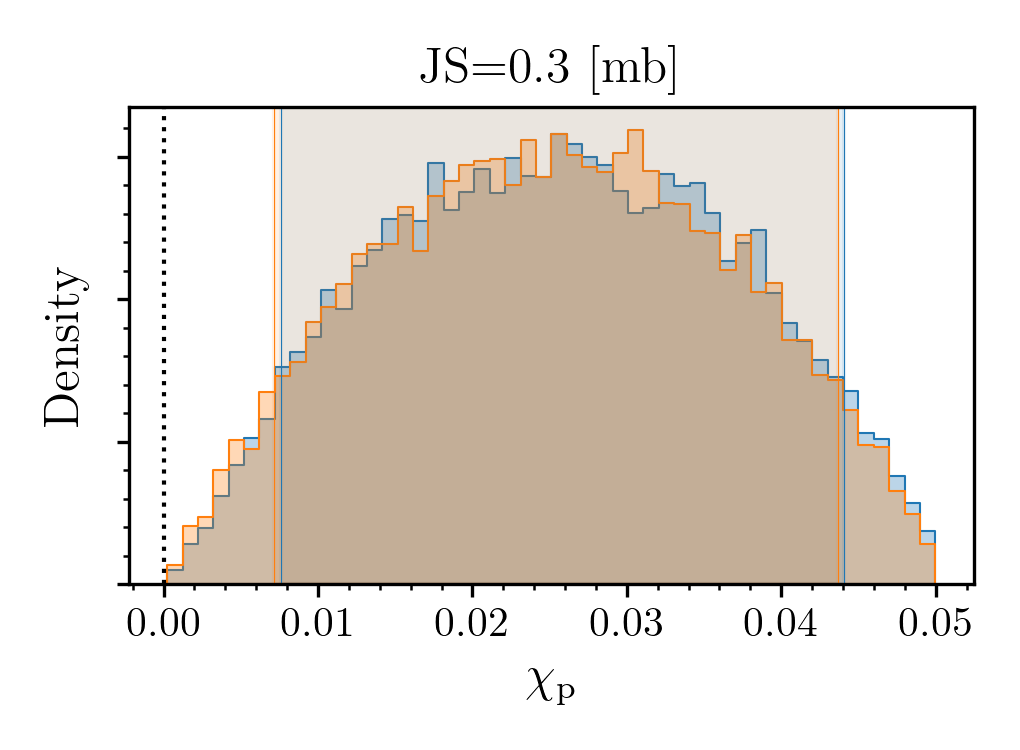}}
\subfigure{\includegraphics[width=.22\textwidth]{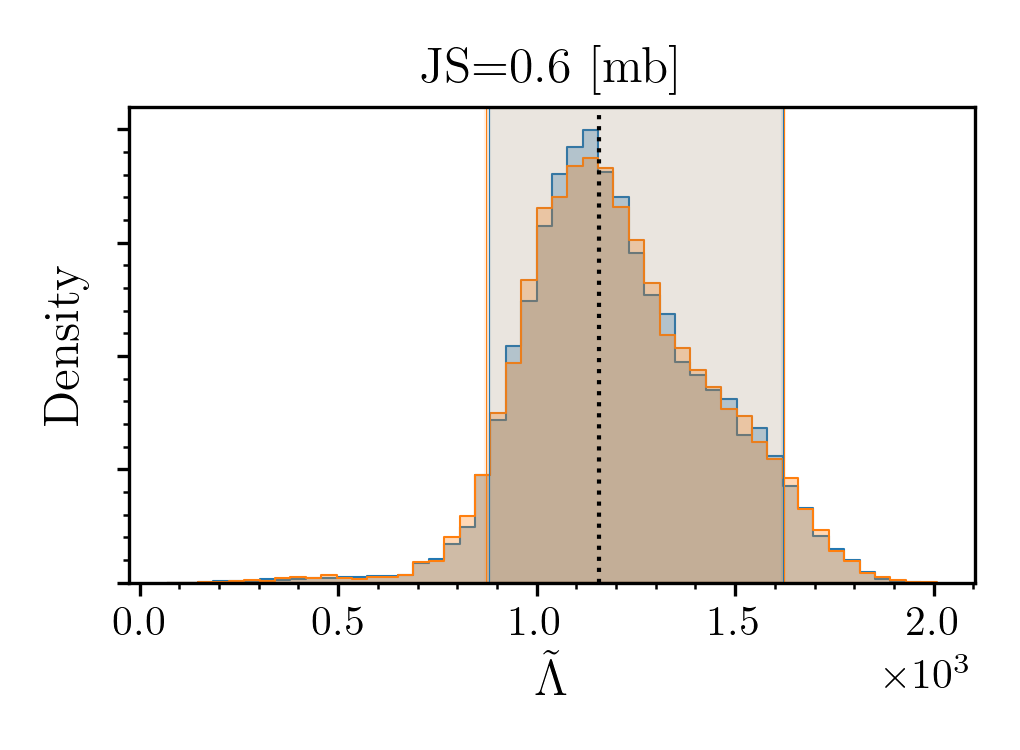}}
\subfigure{\includegraphics[width=.22\textwidth]{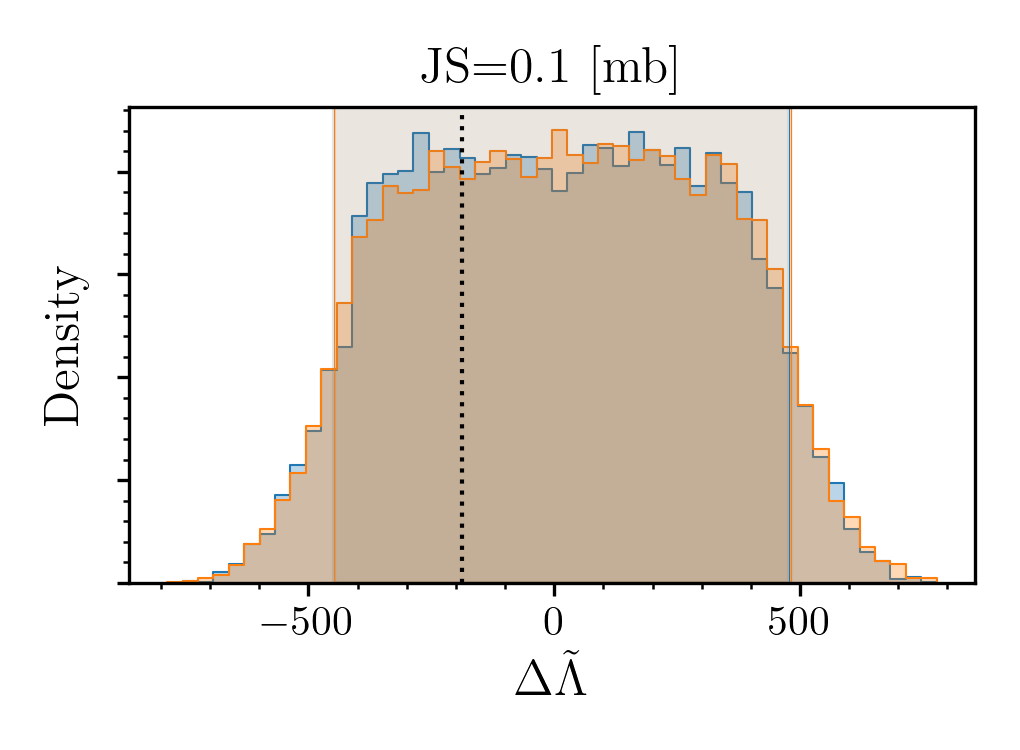}}
\subfigure{\includegraphics[width=.22\textwidth]{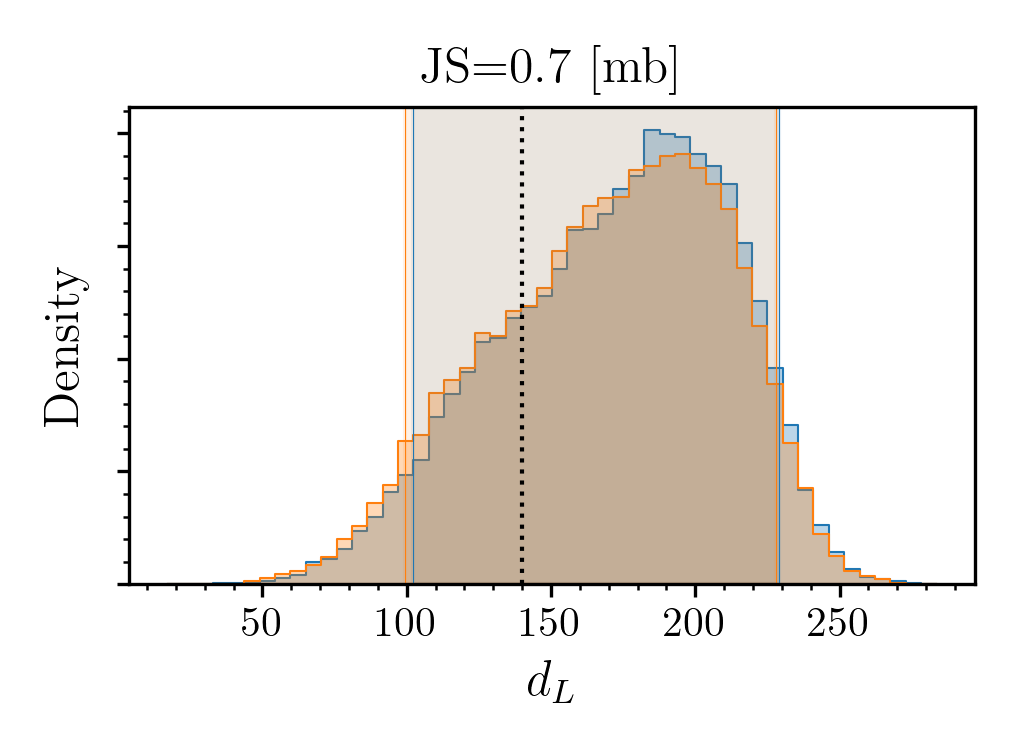}}
\subfigure{\includegraphics[width=.22\textwidth]{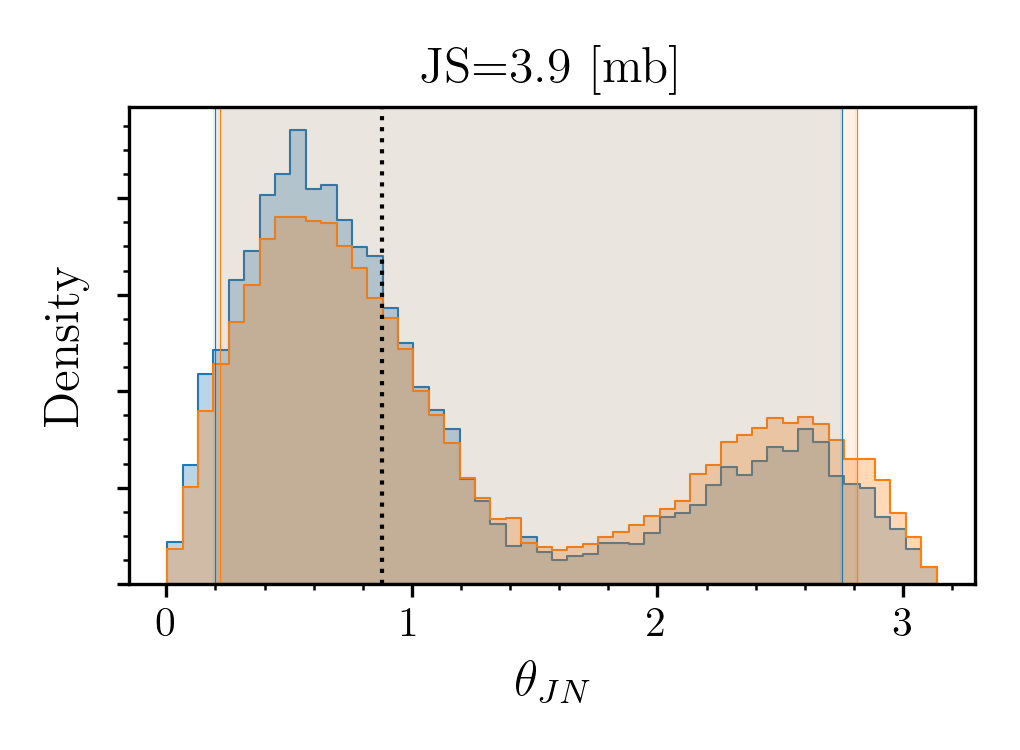}}
\caption{Histograms of selected posteriors from the \dynesty (blue) and \bilbyMCMC (orange) analyses of the fiducial BNS A. Configurations and summary statistics are given in \cref{tab:all}. Vertical lines mark the edges of the 90\% credible interval for each sampler and a black dotted lines marks the value used to simulate the data.
Note that we do not expect the posteriors to peak at the simulation values due to the influence of the simulated noise and the Bayesian prior.
The largest JS-divergence reported across all parameters occurs for the inclination, $\theta_{JN}$, above the threshold of \SI{2}{mb} (see \cref{sec:js}).
}
\label{fig:bns}
\end{figure}

As with the BBH case, the Bayesian evidence estimates (see \cref{tab:all} for the signal vs. noise Bayes factor) disagree. Again, we conclude this is due to the known bias in the parallel-tempered evidence estimate.
The posterior distributions from the \dynesty and \bilbyMCMC are statistically identical, except for the inclination parameter $\theta_{JN}$.
In \cref{fig:bns}, we reproduce histograms for selected parameters of typical astrophysical inference visually demonstrating the agreement and inclination difference.
The cause of the difference in inferred inclination is not yet fully understood, but we note that the difference is only marginally above our threshold for statistically identical.
Comparing individual re-analyses between the two samplers, the difference persists suggesting it is systematic and not a random fluctuation.
We note this is an instance where the posterior is bimodal and speculate this could be a symptom of the \dynesty nested sampling failing to fully explore both modes.
However, the difference is sufficiently small for us to conclude the underlying conclusions about the source (i.e. the 90\% credible intervals) are robust, while the posterior shape is subject to some sampling error (from one or both samplers).

Due to the larger SNR of the fiducial BNS, and the increase in dimension of the prior, the efficiency of both the \dynesty and \bilbyMCMC samplers is reduced compared to that of the fiducial BBH.
The ratio of efficiencies is also increased: the \dynesty sampler is $\sim30$ times more efficient in this case.
As with the BBH analyses, this efficiency does not directly translate into wall-time savings due to the different parallelisation approaches.
However, the efficiency is significant. In future work, we aim to improve the choice of parameterisation and proposals to improve the efficiency of the \bilbyMCMC sampler.

\section{Summary}
\label{sec:conclusion}

We introduce \bilbyMCMC, a parallel-tempered ensemble sampler with problem-specific and machine-learning based proposals.
The \bilbyMCMC sampler is the first MCMC sampler implemented in the \bilby \citep{Bilby1} inference package with demonstrated performance for analysing CBC events observed by ground-based gravitational-wave detectors.
We demonstrate, using both comparisons to known results and cross-sampler comparisons, that the posterior samples are unbiased.
Compared to the \dynesty nested sampling algorithm, \bilbyMCMC suffers a known bias in its estimation of the Bayesian evidence when the number of parallel-tempered chains, \ntemps, is small.
Increasing \ntemps reduces the bias, but at the cost of posterior-sampling efficiency.
We introduce a method to resample from the tempered chains, recovering some of this inefficiency, but find it provides little improvement for typical CBC inference problems.
We conclude that \bilbyMCMC is ideal for problems in which only the posterior distribution is of interest, but that nested sampling approaches should be preferred when evidence calculations are required.
This makes \bilbyMCMC unsuitable for model-comparison via a Bayes factor \citep{mackay2003information}.
Instead, one may wish to develop a hyper-model where the model is treated as a random variable (see e.g. the Reverse Jump Markov Chain Monte Carlo approach described in \citep{2007PhRvD..76h3006C}).

\bilbyMCMC can be trivially and asynchronously parallelised.
This enables it to be configured to leverage High-Throughput Computing environments to reduce the wall-time.
That the parallelisation is asynchronous makes it ideal for utilising non-interacting distributed computing  such as the Open Science Grid \citep{osg07, osg09}.
By comparison, nested sampling approaches can be parallelised solely through the use of multiprocessing.
\citet{pbilby} demonstrated massive scaling of the \dynesty \citep{dynesty} sampler to many hundreds of cores;
\bilbyMCMC cannot similarly be scaled due to the fundamental limit of the burn-in inefficiency.
However, the \citet{pbilby} approach requires synchronised access to a High-Performance Computing environment in which the communication times between cores is rapid.

\bilbyMCMC provides the user access to a modular library of proposal distributions which can be chained together.
The choice of parameterisation and proposals has a significant effect on the efficiency of the sampler.
We anticipate further development in both these aspects will improve the sampler efficiency resulting in reduced wall-time.
Users adapting \bilbyMCMC to other astrophysical inference problems can define their own sets of proposal distributions and easily implement new problem-specific proposals by sub-classing the existing software.

\bibliographystyle{mnras}
\bibliography{bibliography}

\section{Data Availability}
No new data were generated or analysed in support of this research.
The scripts used to perform all verification checks and additional figures are available from \href{https://git.ligo.org/gregory.ashton/bilby_mcmc_validation}{git.ligo.org/gregory.ashton/bilby\_mcmc\_validation}.

\section{Acknowledgements}
We thank Michael Williams, John Veitch, Ben Farr, and Moritz H\"ubner for useful comments during the development of this work.
We thank Will Farr, who's review of this work led to several improvements and clarifications in our exposition.
CT acknowledges support of the National Science Foundation, and the LIGO Laboratory.
We are grateful for computational resources provided by Cardiff University, and funded by an STFC grant ST/I006285/1 supporting UK Involvement in the Operation of Advanced LIGO.
We are also grateful to computing resourced provided by the LIGO Laboratory computing clusters at California Institute of Technology and LIGO Hanford Observatory supported by National Science Foundation Grants PHY-0757058 and PHY-0823459.
 This work makes use of the
\texttt{scipy} \citep{scipy:2020},
\texttt{numpy} \citep{oliphant2006guide, van2011numpy, harris2020array},
and \texttt{pesummary} \citep{Hoy:2020vys}
packages for data analysis and visualisation.

\appendix

\section{The JS-divergence criteria}
\label{sec:js}
\begin{figure}
    \centering
    \includegraphics{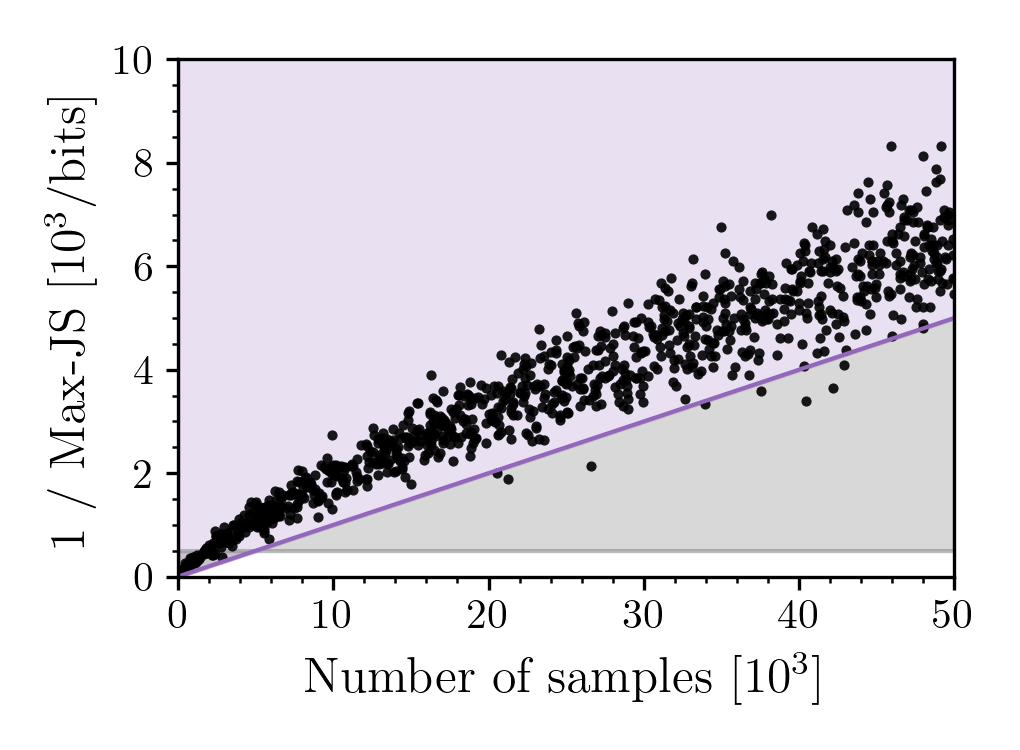}
    \caption{The maximum JSD for 1000 pairs of posteriors drawn form a 15-dimensional unimodal Gaussian distribution.
    We vary the number of samples drawn in each simulation.
    The horizontal gray line indicates the \SI{2}{mb} threshold established in \citet{Bilby2}.
    The purple curve is the new threshold given in \cref{eqn:JS}.
    }
    \label{fig:js}
\end{figure}

As described in \cite{Bilby2}, we use the one-dimensional JSD (maximised over all dimensions) to quantify the agreement between sets of posterior samples.
In that work, a threshold of \SI{2}{mb} was establish for the maximum JSD\footnote{The maximum over the set of sampled parameters}: above this value, the differences between posteriors where deemed statistically significant.
Here, we extend that analysis.
We simulate pairs of posterior samples from the 15-dimensional unimodal Gaussian distribution (cf.~\cref{sec:unimodal}) varying the number of samples drawn in each case.
We find a strong correlation between the number of samples and the inverse of the maximum JSD (\cref{fig:js}).
This demonstrates that, while appropriate for sample sizes of a few thousand, the original threshold is overly conservative for small samples sized and too liberal for larger sample sizes.

To better capture the correlations observed in the simulated data, we introduce a new threshold:
\begin{equation}
    \textrm{maximum JSD} \le \frac{10}{\neffsamples}\,
    \label{eqn:JS}
\end{equation}
This threshold is demonstrated in \cref{fig:js} as the purple shaded region.

For the simulated 15-D system, we see maximum JSD values as large as \cref{eqn:JS} a few times in the 1000 simulations.
This threshold falsely identifies statistical differences between the sets of posterior samples in our simulation as a rate of $\sim 0.1\%$.
In this sense, it can be used as a conservative bound: if the maximum JSD between samplers is found to be larger than the prediction of \cref{eqn:JS}, this highlights an area of concern warranting further study.

We note that a better fit to the lower-bound on the inverse maximum JSD could be found (e.g. by a probability-of-failure based rule), but \cref{eqn:JS} is easy to remember and hence provides a good rule of thumb.

\bsp
\label{lastpage}
\end{document}